\documentclass{aa}

\usepackage{graphicx}
\usepackage[varg]{txfonts}
\usepackage{natbib}
\usepackage[pdftex,colorlinks=true,linkcolor=blue,citecolor=blue,urlcolor=blue]{hyperref}
\usepackage{amsmath}
\usepackage{bbold}
\usepackage{tikz}

\def\i{\mathrm{i}}

\def\expo#1{\mathrm{e}^{#1}}
\def\d{\mathrm{d}}
\def\O#1{\mathcal{O}\left(#1\right)}
\def\Z{\mathbb{Z}}
\def\H{\mathcal{H}}
\def\K{\mathcal{K}}

\def\t{^\mathrm{T}}

\def\primet{^\mathrm{\prime T}}

\def\mean#1{\overline{#1}}
\def\varmean#1{\left\langle#1\right\rangle}
\def\bracktwolines#1#2{#1\vphantom{#2}\right.\nonumber\\&\left.\vphantom{#1}#2}
\def\kepmodelURL{\url{https://gitlab.unige.ch/jean-baptiste.delisle/kepmodel}}

\DeclareMathOperator{\diag}{diag}

\DeclareMathOperator{\sinc}{sinc}

\def\hadprod{\!*\!}

\begin{document}

\title{Analytical determination of orbital elements using Fourier analysis}
\subtitle{II. Gaia astrometry and its combination with radial velocities%
  \thanks{We provide an open-source implementation of the methods presented in this article:
    the kepmodel python package, installable with pip or conda. The sources are available at
    \kepmodelURL.}}
\author{J.-B. Delisle
  \and D. Ségransan
}
\institute{Département d'astronomie, Université de Genève,
  chemin Pegasi 51, 1290 Versoix, Switzerland\\
  \email{jean-baptiste.delisle@unige.ch}
}
\date{\today}

\abstract{
  The ESA global astrometry space mission Gaia has been monitoring
  the position of a billion stars since 2014.
  The analysis of such a massive dataset is challenging in terms of the data processing involved.
  In particular, the blind detection and characterization of
  single or multiple companions to stars (planets, brown dwarfs, or stars)
  using Gaia astrometry requires highly efficient algorithms.
  In this article, we present a set of analytical methods to detect and characterize companions
  in scanning space astrometric time series
  as well as via a combination of astrometric and radial velocity time series.
  We propose a general linear periodogram framework
  and we derive analytical formulas for the false alarm probability (FAP) of periodogram peaks.
  Once a significant peak has been identified,
  we provide analytical estimates of all the orbital elements of the companion
  based on the Fourier decomposition of the signal.
  The periodogram, FAP, and orbital elements estimates
  can be computed for the astrometric and radial velocity time series
  separately or in tandem.
  These methods are complementary with more accurate
  and more computationally intensive numerical algorithms
  (e.g., least-squares minimization, Markov chain Monte Carlo, genetic algorithms).
  In particular, our analytical approximations can be used as an initial condition to
  accelerate the convergence of numerical algorithms.
  Our formalism has been partially implemented in the Gaia exoplanet pipeline
  for the third Gaia data release.
  Since the Gaia astrometric time series are not yet publicly available,
  we illustrate our methods on the basis of Hipparcos data,
  together with on-ground CORALIE radial velocities,
  for three targets known to host a companion:
  \object{HD~223636} (HIP~117622), \object{HD~17289} (HIP~12726), and \object{HD~3277} (HIP~2790).
}


\maketitle

\section{Introduction}
\label{sec:introduction}

The Gaia mission started its nominal operations in 2014 to monitor the positions of a billion stars.
The amount of data generated by the spacecraft is massive and their calibration and analysis
represent a vast processing challenge.
In this context, the computing time dedicated to the analysis of each target
is necessarily limited and efficient analysis methods have to be developed.
In this article, we present efficient analytical methods
for detecting and characterizing companions
(planets, brown dwarfs, or stars)
in astrometric time series
as well as via a combination of astrometric and radial velocity time series.
This study is part of an effort
to improve the Gaia exoplanet detection pipeline
and, more specifically, the MIKS-GA\footnote{MIKS-GA: multiple independent Keplerian solution using a Genetic algorithm}
algorithms \citep{holl_2022_gaia}.
The methods presented here have been partially integrated in MIKS-GA for
the third Gaia data release (DR3).

The least-squares or Lomb-Scargle (LS) periodogram \citep{lomb_1976_leastsquares,scargle_1982_studies}
and subsequent generalizations \citep[e.g.,][]{ferraz-mello_1981_estimation,zechmeister_2009_generalised}
are powerful tools in the search for periodicities in uneven time series
and widely used in many branches of astronomy.
In the context of two-dimensional (2D) astrometry
(i.e., when both sky coordinates are measured at once, with a similar level of precision),
the LS periodogram has also been used
to search for periodicities on each coordinate independently \citep[e.g.,][]{black_1982_detection}
or jointly \citep[e.g.,][]{catanzarite_2006_astrometric}.
More sophisticated periodograms have also been developed in the case of 2D astrometry,
for instance, the phase distance correlation periodogram \citep{zucker_2019_detection}.
However, to our knowledge, these approaches have not been extended to the case of scanning space astrometric missions,
such as Hipparcos and Gaia.
Here, we propose a general linear periodogram framework to efficiently locate
the periods of candidate companions among scanning space astrometric time series alone and
among astrometric and radial velocity time series combined.
We additionally derive an analytical false alarm probability (FAP) criterion
to assess the significance of the candidates.
This periodogram and FAP framework builds on the previous works of
\citet{baluev_2008_assessing} and \citet{delisle_2020_efficient},
which were primarily focused on the analysis of radial velocity time series.

Once a companion is detected and its period is known,
the next step is to determine the complete set of parameters describing its orbit.
This is usually achieved using numerical methods,
such as least-squares minimization (or likelihood maximization) algorithms,
or Markov chain Monte-Carlo.
However these numerical methods can be expensive in terms of computing time,
especially if the initialization of the parameters is far from the solution.
Here, we propose an analytical method that can efficiently determine approximate orbital elements,
which may be used as the initial conditions for numerical methods.
It relies on the Fourier decomposition of the time series.
Indeed, for an eccentric orbit, the signature of the companion
has some power at the orbital period, but also at the harmonics of the orbital period.
In particular, the amplitude at the first harmonics is approximately proportional to the eccentricity.
We used such properties to derive all the orbital parameters from a linear least-square fit with
sinusoids at the orbital period and its first harmonics.
Similar methods have been developed in the context of astrometry with
the cancelled space interferometry mission (SIM) \citep{konacki_2002_frequency}
and radial velocities \citep{correia_2008_determination,delisle_2016_analytical}.
In this work, we extend the method proposed in \citet{delisle_2016_analytical}
to the case of scanning space astrometry as well as
to the case where astrometry and radial velocities are combined.

This article is organized as follows.
In Sect.~\ref{sec:astrometry}, we present the scanning space astrometric method
and the expected astrometric signature of an unseen companion orbiting a star.
Then, we define in Sect.~\ref{sec:periofap} a general linear periodogram
for an astrometric time series and its combination with radial velocities
and we derive analytical approximations of the corresponding FAP.
We obtain in Sect.~\ref{sec:orbparam} analytical formulas to derive an approximation of
all the orbital parameters once the period is known.
We illustrate our methods in Sect.~\ref{sec:applications}
to reanalyze the astrometric and radial velocity time series of
three stars known to host companions.
While our methods have been developed in view of analyzing Gaia data,
we had to fall back to illustrating them on the basis of Hipparcos data,
since Gaia astrometric time series will not be publicly available before DR4.
Finally, we discuss in Sect.~\ref{sec:discussion} the possible uses of these methods.

\section{Scanning space astrometry and Keplerian motion}
\label{sec:astrometry}

\subsection{Coordinates and scan-angle}
\label{sec:scanangle}

\begin{figure}
  \centering
  \begin{tikzpicture}
    \pgfmathsetmacro{\th}{55}
    \pgfmathsetmacro{\scale}{3.5}
    \pgfmathsetmacro{\lw}{1pt}
    \coordinate (O) at (0,0);
    \coordinate (d) at (0,\scale);
    \coordinate (a) at (-\scale,0);
    \coordinate (AL) at  ({-\scale*sin(\th)},{\scale*cos(\th)});
    \coordinate (AC) at  ({\scale*cos(\th)},{\scale*sin(\th)});
    \draw[line width=\lw,-stealth] (O) -- (d) node[above] {$\delta$};
    \draw[line width=\lw,-stealth] (O) -- (a) node[left] {$\alpha$};
    \draw[line width=\lw,->] (O) -- (AL) node[left] {AL};
    \draw[line width=\lw,->] (O) -- (AC) node[above] {AC};
    \draw[line width=\lw,->] (O)+(90:{0.2*\scale}) arc (90:{90+\th}:{0.2*\scale}) node[midway,above] {$\theta$};
    \draw[line width=\lw,->] (O)+({90+\th}:{0.3*\scale}) arc ({90+\th}:{180}:{0.3*\scale}) node[midway,left] {$\psi$};
  \end{tikzpicture}
  \caption{Scan angle convention for Gaia ($\theta$) and Hipparcos ($\psi$).}
  \label{fig:scanangle}
\end{figure}
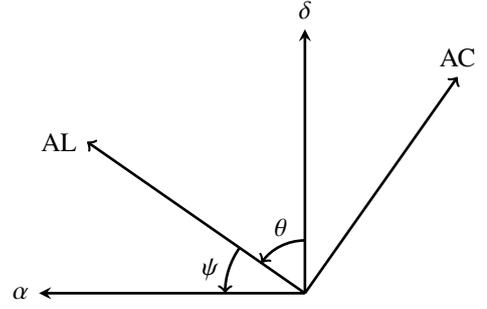

Scanning space astrometric missions such as Hipparcos or Gaia do not directly measure
the instantaneous declination ($\delta$) and right-ascension ($\alpha$)
of a star in the ICRS frame.
They instead measure the abscissa, $s_\mathrm{AL}$, of the star along a scan direction with great precision,
while the perpendicular coordinate, $s_\mathrm{AC}$, is determined with a much lower precision.
The measured coordinates $(s_\mathrm{AL}$, $s_\mathrm{AC})$ depend on the scanning direction
which is represented with a different convention between Gaia and Hipparcos (see Fig.~\ref{fig:scanangle}).
The Gaia scan angle $\theta$ is
the angle between the north (increasing $\delta$)
and the along-scan direction (increasing $s_\mathrm{AL}$),
while the Hipparcos scan angle $\psi$ is
the angle between the along-scan direction
and the East (increasing $\alpha$).
We thus have:
\begin{align}
  \psi & = \frac{\pi}{2} - \theta,\nonumber                                           \\
  s_\mathrm{AL}
       & = \delta  \cos\theta + \alpha^* \sin\theta + \varpi \Pi_\mathrm{AL}\nonumber \\
       & = \delta \sin\psi + \alpha^* \cos\psi + \varpi \Pi_\mathrm{AL},\nonumber     \\
  s_\mathrm{AC}
       & = \delta  \sin\theta - \alpha^* \cos\theta + \varpi \Pi_\mathrm{AC}\nonumber \\
       & = \delta  \cos\psi - \alpha^* \sin\psi + \varpi \Pi_\mathrm{AC},
\end{align}
where
$\alpha^* = \alpha\cos\delta$ is the modified right-ascension,
$\varpi$ is the parallax,
and $\Pi_\mathrm{AL}$, $\Pi_\mathrm{AC}$ are the parallax factors.
Across scan observations can actually be treated
the same way as along scan observations by defining
$\theta'=\theta-\pi/2$ and $\psi' = \psi + \pi/2$.
In the following, we assume that all the observations (Gaia or Hipparcos, along or across scan)
are converted to follow the Gaia along-scan convention
\begin{equation}
  \label{eq:xastro}
  s = \delta  \cos\theta + \alpha^* \sin\theta + \varpi \Pi.
\end{equation}

The instantaneous coordinates of the star ($\delta, \alpha^*$) evolve
due to the proper motion of the system barycenter
and to potential companions.
The proper motion can be expressed as a polynomial expansion over time,
so the coordinates follow
\begin{align}
  \label{eq:alphadelta}
  \delta(t)   & = \delta_0 + \sum_{k=1}^{k_\mathrm{max}} \mu_{\delta}^{(k)} t^k + \delta_K(t),\nonumber \\
  \alpha^*(t) & = \alpha_0^* + \sum_{k=1}^{k_\mathrm{max}} \mu_{\alpha^*}^{(k)} t^k + \alpha_K(t),
\end{align}
where
$(\delta_0, \alpha_0^*)$ is the position of the system barycenter at the reference epoch ($t=0$),
${k_\mathrm{max}}$ is the degree of the expansion of the proper motion,
$\mu_\delta^{(k)}$, $\mu_{\alpha^*}^{(k)}$ are the coefficients of this expansion,
and ($\delta_K$, $\alpha_K$) is the Keplerian part of the signal (due to companions).
In most cases, only a linear proper motion is accounted for ($k_\mathrm{max}=1$),
and we simplify the notations as
$\mu_\delta = \mu_\delta^{(1)}$, $\mu_{\alpha^*} = \mu_{\alpha^*}^{(1)}$

\subsection{Keplerian astrometric signal}
\label{sec:astrokep}

Here, we consider a star that is orbited by a single companion, assuming that the orbit is Keplerian.
This Keplerian motion is given by
\begin{align}
  \label{eq:keplerianTI}
  \delta_K & = A x + F y,\nonumber \\
  \alpha_K & = B x + G y,
\end{align}
where
\begin{align}
  \label{eq:xyfromeE}
  x & = \cos E - e,\nonumber \\
  y & = \sqrt{1-e^2} \sin E,
\end{align}
with $e$ as the eccentricity, $E$ the eccentric anomaly, and
$A$, $B$, $F$, $G$ the Thiele-Innes coefficients
\begin{align}
  \label{eq:ThieleInnes}
  A & = a_s (\cos\omega\cos\Omega - \sin\omega\sin\Omega\cos i),\nonumber  \\
  B & = a_s (\cos\omega\sin\Omega + \sin\omega\cos\Omega\cos i),\nonumber  \\
  F & = a_s (-\sin\omega\cos\Omega - \cos\omega\sin\Omega\cos i),\nonumber \\
  G & = a_s (-\sin\omega\sin\Omega + \cos\omega\cos\Omega\cos i),
\end{align}
where $\omega$, $\Omega$, and $i$ are the argument of periastron,
longitude of ascending node, and inclination
of the orbit, respectively,
and $a_s$ is the semi-major axis of the orbit of the star around the center of mass of the system,
expressed in mas.
The eccentric anomaly is implicitly expressed as a function of time
through the Kepler equation
\begin{equation}
  \label{eq:Keplereq}
  E - e \sin E = M = M_0 + 2\pi \frac{t}{P},
\end{equation}
where $M$ is the mean anomaly, $M_0$ is the mean anomaly at the reference epoch, and $P$ is the orbital period.

\subsection{Fourier decomposition of a Keplerian astrometric signal}
\label{sec:fourier}

\begin{figure}
  \centering
  \includegraphics[width=\linewidth]{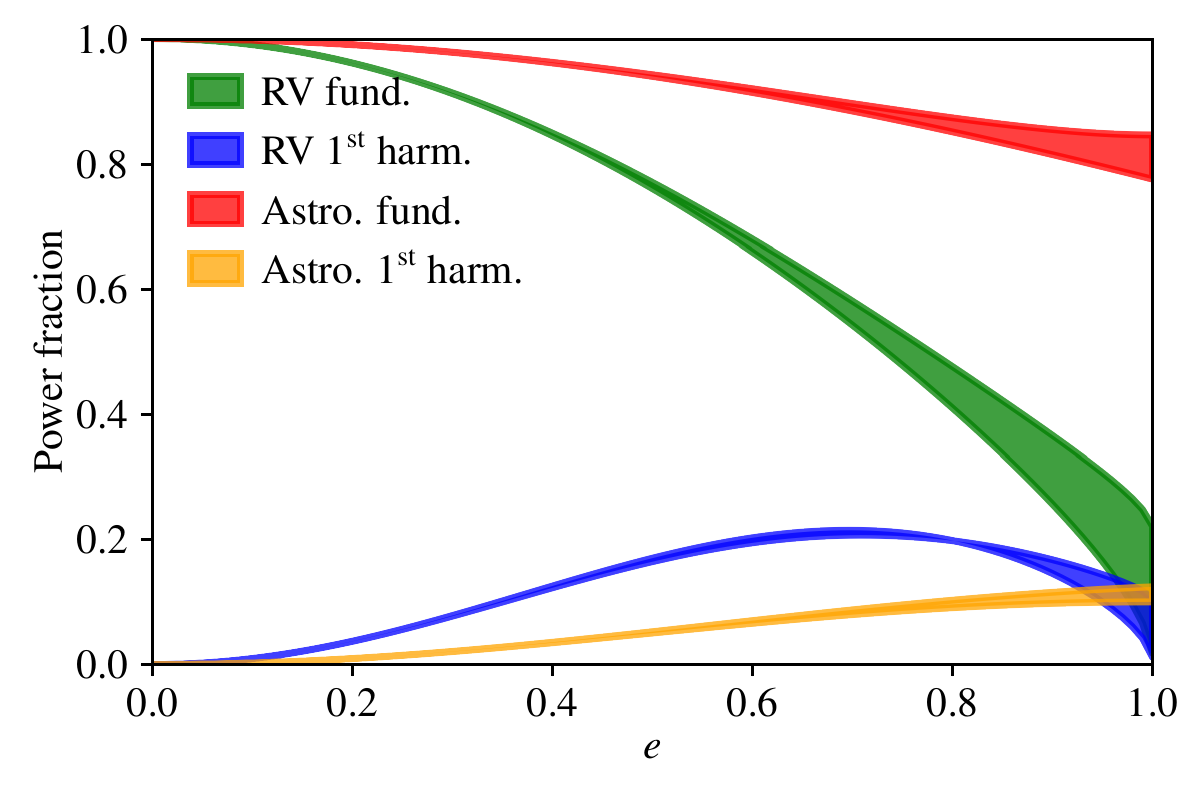}
  \caption{Fraction of the signal power in the fundamental and in the first harmonics for a radial velocity time series and an astrometric time series.}
  \label{fig:power_fraction}
\end{figure}

The Keplerian part of the astrometric signal is $P$-periodic and can be decomposed as a discrete Fourier series
\begin{align}
  \label{eq:defdkak}
  \delta_K & = \sum_{k\in\Z} d_k \expo{\i knt},\nonumber \\
  \alpha_K & = \sum_{k\in\Z} a_k \expo{\i knt},
\end{align}
with $n = 2\pi/P$ as the planet mean-motion.
We introduce the complex coordinate,
\begin{equation}
  \zeta = x + \i y = \cos E - e + \i \sqrt{1-e^2}\sin E,
\end{equation}
and the complex coefficients,
\begin{align}
  P_\delta \expo{\i \phi_\delta} & = \frac{A - \i F}{2},\nonumber \\
  P_\alpha \expo{\i \phi_\alpha} & = \frac{B - \i G}{2},
\end{align}
such that
\begin{align}
  \delta_K & = P_\delta\left( \expo{\i \phi_\delta} \zeta + \expo{-\i \phi_\delta} \bar{\zeta}\right),\nonumber \\
  \alpha_K & = P_\alpha\left( \expo{\i \phi_\alpha} \zeta + \expo{-\i \phi_\alpha} \bar{\zeta}\right).
\end{align}
We then write the Fourier decomposition of $\zeta$ in the following form:
\begin{equation}
  \zeta(t) = \sum_{k\in\Z} \zeta_k \expo{\i k M} = \sum_{k\in\Z} \zeta_k \expo{\i k M_0} \expo{\i k n t},
\end{equation}
where
\begin{equation}
  \label{eq:defzetak}
  \zeta_k = \frac{1}{2\pi} \int_0^{2\pi} \zeta \expo{-\i k M} \d M
\end{equation}
are functions of the eccentricity only.
The coefficients $d_k$ and $a_k$ are then easily expressed as functions of the coefficients $\zeta_k$
\begin{align}
  \label{eq:dkak}
  d_k & = P_\delta \expo{\i kM_0} \left( \expo{\i \phi_\delta} \zeta_k + \expo{-\i \phi_\delta} \zeta_{-k}\right),\nonumber \\
  a_k & = P_\alpha \expo{\i kM_0} \left( \expo{\i \phi_\alpha} \zeta_k + \expo{-\i \phi_\alpha} \zeta_{-k}\right).
\end{align}
The coefficients $\zeta_k$ can be computed numerically (see Appendix~\ref{sec:power_fraction}).
They can also be computed analytically as power series of the eccentricity.
We go on to rewrite Eq.~(\ref{eq:defzetak}) as an integral over the eccentric anomaly, $E$,
\begin{align}
  \label{eq:zetakEA}
  \zeta_k = \frac{1}{2\pi} \int_0^{2\pi} & \left(\cos E - e + \i \sqrt{1-e^2}\sin E\right)\nonumber \\
                                         & \times \expo{-\i k (E-e\sin E)} (1-e\cos E) \d E.
\end{align}
This expression can then straightforwardly be developed in power series of the eccentricity
and integrated over $E$.
We obtain for the first terms:
\begin{align}
  \label{eq:zetak}
  \zeta_0    & = -\frac{3}{2} e,\nonumber                            \\
  \zeta_1    & = 1 - \frac{1}{2} e^2 + \O{e^4},\nonumber             \\
  \zeta_{-1} & = \frac{1}{8} e^2 + \O{e^4},\nonumber                 \\
  \zeta_2    & = \frac{1}{2} e - \frac{3}{8} e^3 + \O{e^5},\nonumber \\
  \zeta_{-2} & = \frac{1}{24} e^3 + \O{e^5}.
\end{align}
At low eccentricity, the leading Fourier coefficients of $\delta$ and $\alpha$
are found at the orbital period.
The amplitude of the first harmonics (half the orbital period) terms
is proportional to the eccentricity.
Higher order harmonics have a much lower amplitude (higher powers of the eccentricity).
In the following, we aim to detect the presence of a companion using a linear periodogram approach
where we approximate the Keplerian signal with the fundamental frequency only.
Then, once the companion is detected and its orbital period is known, we
estimate the remaining parameters by approximating the Keplerian signal using the fundamental and the first harmonics.
These two approximations remain valid for low to moderate eccentricity.
In order to assess the domain of validity of this approach, we estimated
the fraction of the signal power that lies in the fundamental and in the first harmonics.
We show, in Fig.~\ref{fig:power_fraction}, these power fractions as a function of the eccentricity.
As a comparison, we also provide the same fractions in the case of a radial velocity Keplerian signal.
The details of these derivations are presented in Appendix~\ref{sec:power_fraction}.
As expected, we see in Fig.~\ref{fig:power_fraction} that at low eccentricity,
the power of the signal is concentrated in the fundamental frequency both for a radial velocity signal and an astrometric signal.
However, the behavior of both kinds of signal differs at higher eccentricity.
For the radial velocities, the power fraction in the fundamental
drops to zero when the eccentricity reaches 1.
Therefore, for a radial velocity signal,
the linear periodogram approach (with a single frequency)
has a significant loss of sensitivity for high eccentricity signals
\citep[see also][]{baluev_2015_keplerian}.
To circumvent this issue, more complex periodogram definitions have been proposed in the literature
\citep[e.g.,][]{baluev_2013_vonmises,baluev_2015_keplerian,zucker_2018_detection}.
In the case of astrometry, the power fraction in the fundamental also decreases with eccentricity, but it always remains above about 80\,\%.
Therefore, the linear periodogram approach
remains fairly sensitive even for highly eccentric orbits,
and a more complex periodogram approach would probably be less beneficial
than in the radial velocity case.

While a high power fraction in the fundamental improves the detection sensitivity
of the linear periodogram, it also reduces the ability to characterize the
companion's eccentricity and argument of periastron.
Indeed, the fundamental frequency alone does not contain sufficient information to derive
these parameters.
Therefore, one can only derive the eccentricity and argument of periastron
with good precision for a signal that has significant power in the harmonics.
This applies in particular to our analytical method (Sect.~\ref{sec:orbparam})
but, more generally, independently of the method,
we expect the eccentricity and argument of periastron to be
characterized with less precision using astrometry than using radial velocities,
for a similar signal to noise ratio (S/N).

\section{Periodogram and false alarm probability}
\label{sec:periofap}

\subsection{General linear periodogram}
\label{sec:periodef}

Here, we adopt a periodogram approach to detect the signature of a companion and estimate its orbital period.
Following \citet{baluev_2008_assessing,delisle_2020_efficient},
we define a general linear periodogram by comparing
the $\chi^2$ of the residuals of a linear base model, $\H$, with $p$ parameters
and enlarged linear model, $\K$, of $p+d$ parameters, further parameterized by the angular frequency, $\nu$.
The linear base model, $\H$, follows:
\begin{equation}
  \label{eq:mh}
  \H\ :\quad m_\H(\beta_\H) = \varphi_\H\beta_\H,
\end{equation}
where $\beta_\H$ is the vector of size $p$ of the model parameters,
$\varphi_\H$ is a $n\times p$ matrix, and $n$ is the number of points in the time series.
Typically, this base model includes five linear parameters,
including the position, the proper motion, and the parallax
(see Eqs.~(\ref{eq:xastro}),~(\ref{eq:alphadelta})):
\begin{align}
  \label{eq:basemodastro}
  \beta_\H   & = (\delta_0, \alpha_0^*, \mu_\delta, \mu_\alpha, \pi),\nonumber \\
  \varphi_\H & =
  \begin{pmatrix}
    \cos\theta_1 & \sin\theta_1 & t_1 \cos\theta_1 & t_1 \sin\theta_1 & \Pi_1  \\
    \cos\theta_2 & \sin\theta_2 & t_2 \cos\theta_2 & t_2 \sin\theta_2 & \Pi_2  \\
    \vdots       & \vdots       & \vdots           & \vdots           & \vdots \\
    \cos\theta_n & \sin\theta_n & t_n \cos\theta_n & t_n \sin\theta_n & \Pi_n
  \end{pmatrix},
\end{align}
but any additional linear parameters might be included.
For instance, a quadratic proper motion can be accounted for by adding the columns
$t^2 \cos\theta, t^2 \sin\theta$ in $\varphi_\H$.

The enlarged model $\K(\nu)$ follows
\begin{equation}
  \label{eq:mk}
  \K(\nu)\ :\quad m_\K(\nu, \beta_\K) = \varphi_\K(\nu)\beta_\K,
\end{equation}
where $\beta_\K = (\beta_\H, \beta)$ is the vector of the $p+d$ linear parameters,
and $\varphi_\K(\nu)$ is the $n\times(p+d)$ matrix whose $p$ first columns are the columns of $\varphi_\H$,
and whose $d$ last columns are functions of the angular frequency, $\nu$.

In the case of radial velocities \citep[see][]{baluev_2008_assessing,delisle_2020_efficient},
we would typically have $d=2$, with the additional columns
of $\cos\nu t$ and $\sin\nu t$,
to account for any possible phase of the potential companion.
This definition implicitly assumes that the signature of the companion
is dominated by the fundamental frequency (see Sect.~\ref{sec:fourier}).
In the astrometric case, the same approximation is achieved
with the $d=4$ additional columns:
$\cos\theta\cos\nu t$, $\sin\theta\cos\nu t$,
$\cos\theta\sin\nu t$, and $\sin\theta\sin\nu t$.
These four terms allow us to account for any possible phase and for the projection
of the orbit over both axes (see Eqs.~(\ref{eq:xastro}),~(\ref{eq:alphadelta})).
Thus, we typically use
\begin{align}
  \label{eq:linmodnu}
  \beta_\K        & = \left(\beta_\H, \beta_{\delta, c}, \beta_{\alpha, c}, \beta_{\delta, s}, \beta_{\alpha, s}\right),                    \\
  \varphi_\K(\nu) & = \left(\varphi_\H, \cos\theta\cos\nu t, \sin\theta\cos\nu t, \cos\theta\sin\nu t, \sin\theta\sin\nu t\right),\nonumber
\end{align}
where $\beta_\K$ is the vector of the $p+4$ parameters,
and $\varphi_\K$ is the $(n\times (p+4))$ matrix obtained by horizontal concatenation of $\varphi_\H$
and the four additional columns.

We denote by $\chi_\H^2$ and $\chi_\K^2(\nu)$ the $\chi^2$ of the residuals
after a linear least squares fit
with a covariance matrix $C$
of the models $\H$ and $\K(\nu)$, respectively:
\begin{align}
  \chi_\H^2      & = (s - \varphi_H\beta_H)^T C^{-1} (s - \varphi_H\beta_H),\nonumber  \\
  \chi_\K^2(\nu) & = (s - \varphi_K(\nu)\beta_K)^T C^{-1} (s - \varphi_K(\nu)\beta_K).
\end{align}
The covariance matrix, $C$, may account for all sources of correlated and uncorrelated noise,
and in particular the instrument systematics.

In the general case, the periodogram is a function $z(\nu) = f(\chi_\H^2, \chi_\K^2(\nu))$.
A general linear periodogram $z$ is thus defined by the models, $\H$ and $\K(\nu)$, and the function $f$.
For instance, the widespread GLS periodogram
\citep[see][]{ferraz-mello_1981_estimation,zechmeister_2009_generalised}
is defined as
\begin{equation}
  \label{eq:GLS}
  z_\mathrm{GLS}(\nu) = \frac{\chi_\H^2 - \chi_\K^2(\nu)}{\chi_\H^2}.
\end{equation}
\citet{baluev_2008_assessing} proposed four alternative definitions of the periodogram.
In the following, we focus on the GLS periodogram since it is more widely used,
but similar derivations can be found in Appendix~\ref{sec:fapcomputation}
for these alternative definitions.

\subsection{False alarm probability}
\label{sec:fap}

\begin{table}
  \caption{False alarm probability for the GLS periodogram (see Eq.~(\ref{eq:GLS})) following the method of \citet{baluev_2008_assessing}.}
  {\renewcommand{\arraystretch}{2}\setlength{\tabcolsep}{3pt}\centering
    \begin{tabular}{l|ll}
      \hline
      \hline
      $d$ & $\mathrm{FAP_{single}}(Z)$                                                                                     & $\tau(Z, \nu_\mathrm{max})$                                    \\
      \hline
      --  & $\displaystyle 1-\frac{\mathrm{B}(Z; d/2, n_\K/2)}{\mathrm{B}(d/2, n_\K/2)}$                                   & $\gamma W \left(1-Z\right)^\frac{n_\K-1}{2} Z^{\frac{d-1}{2}}$ \\
      2   & $\left(1-Z\right)^\frac{n_\K}{2}$                                                                              & $\gamma W\left(1-Z\right)^\frac{n_\K-1}{2}\sqrt{Z}$            \\
      4   & $\left(1+\frac{n_\K}{2}Z\right)\left(1-Z\right)^\frac{n_\K}{2}$                                                & $\gamma W\left(1-Z\right)^\frac{n_\K-1}{2}Z^{\frac{3}{2}}$     \\
      6   & $\left(1+\frac{n_\K}{2}Z+\frac{n_\K}{2}\left(\frac{n_\K}{2}+1\right)Z^2\right)\left(1-Z\right)^\frac{n_\K}{2}$ & $\gamma W\left(1-Z\right)^\frac{n_\K-1}{2}Z^{\frac{5}{2}}$     \\
      \hline
    \end{tabular}}
  \tablefoot{The factor $W$ is the rescaled frequency bandwidth
  defined in Eq.~(\ref{eq:defW}),
  $\mathrm{B}(a, b)$ is the beta function, $\mathrm{B}(x; a, b)$ is the incomplete beta function.
  The factor
  $\gamma = \Gamma\left(\frac{n_\H}{2}\right)/\Gamma\left(\frac{n_\K+1}{2}\right)$,
  where $\Gamma(s)$ is Euler's gamma function,
  can be approximated by $\gamma \sim \left(\frac{n_\H}{2}\right)^{\frac{d-1}{2}}$ for $n_\H \geq 10$. }
  \label{tab:power}
\end{table}

Once the periodogram is computed, it is useful to compute the $p$-value of the highest peak (FAP), defined as:
\begin{equation}
  \mathrm{FAP_{max}}(Z, \nu_\mathrm{max}) = \mathrm{Pr}\{\max_{\nu<\nu_\mathrm{max}} z(\nu) \geqslant Z\ |\ \H\},
\end{equation}
where $Z$ is the value of the maximum peak of the periodogram computed on the data.
In this section, we provide analytical approximations of the FAP
for the GLS periodogram of Eq.~(\ref{eq:GLS}).
Similar approximations for the alternative periodogram definitions and
their precise derivation is provided in Appendix~\ref{sec:fapcomputation}.

The model $\H$ is defined as in Eq.~(\ref{eq:mh}),
where the $n \times p$ matrix $\varphi_\H$ is user-defined,
and the model $\K$ is defined as in Eq.~(\ref{eq:linmodnu}).
The periodogram is computed in the range of frequencies $]0,\nu_\mathrm{max}]$
(i.e., 0 excluded and $\nu_\mathrm{max}$ included).
The FAP is approximated by
\citep[see][]{baluev_2008_assessing}:
\begin{equation}
  \label{eq:FAPmax}
  \mathrm{FAP_{max}}(Z, \nu_\mathrm{max}) \approx 1 - \left(1-\mathrm{FAP_{single}}(Z)\right)\expo{-\tau(Z, \nu_\mathrm{max})},
\end{equation}
where analytical expressions for $\mathrm{FAP_{single}}$ and $\tau(Z, \nu_\mathrm{max})$
are given in Table~\ref{tab:power} ($d=4$).
These expressions depend on the rescaled frequency bandwidth, $W$, defined as
\begin{equation}
  \label{eq:defW}
  W = \frac{\nu_\mathrm{max}}{2\pi} T_\mathrm{eff},
\end{equation}
where $T_\mathrm{eff}$ is the effective time series length.
We provide in Appendix~\ref{sec:fapcomputation} analytical estimates of $T_\mathrm{eff}$
in the general case of correlated noise.
In the simplest case of white noise, that is,
for a diagonal covariance matrix $C = \diag(\sigma^2)$,
the effective time series length can be approximated
by
\begin{equation}
  \label{eq:Teffwhite}
  T_\mathrm{eff} \approx \sqrt{4\pi\left(\mean{t^2}-\mean{t}^2\right)},
\end{equation}
where $\mean{t}$ and $\mean{t^2}$ are weighted means with weights $\sigma_i^2/\sum_j \sigma_j^2$.
This expression is proportional to the weighted standard deviation of the measurement times.

The FAP estimate that we obtain here rely on the correct estimation of the power, $Z$, of the highest peak
in the considered range of angular frequency $]0,\nu_\mathrm{max}]$.
When computing the periodogram, we should thus sample $\nu$
with a sufficiently small step size to determine $Z$ with a good precision.
Moreover, it is also important for the following steps to correctly identify the
period of the highest peak, which also requires for the periodogram to be correctly sampled.
The width of periodogram peaks is typically $2\pi/T_{\mathrm{eff}}$ (in angular frequency).
Each peak should be sampled with a sufficient number of points to obtain a good estimation
of the peak height.
A standard procedure in the literature is to sample the angular frequency linearly
with a step size of $2\pi/n_o T$
where $T$ is the total time span of the observations (which is a proxy for $T_\mathrm{eff}$),
and $n_o \gtrsim 10$ is the oversampling factor.
The maximum angular frequency $\nu_\mathrm{max}$,
or equivalently the minimum period $P_\mathrm{min}$, is user-defined.
By increasing the size of the frequency range,
we increase the probability of obtaining a higher peak by chance (null hypothesis),
which is why the FAP estimate takes into account the maximum frequency, $\nu_\mathrm{max}$,
through the rescaled frequency bandwidth, $W$.

\subsection{Combining astrometric and radial velocity time series}
\label{sec:combperiofap}

We now consider the case where both astrometric and radial velocity time series
are available for a source.
We extend to this case the periodogram and analytical FAP approach described above.
In the following, we assume that the calendars of astrometric and radial velocity observations are independent
and that the noise affecting both time series is independent.

The general linear periodogram framework described above is compatible with heterogeneous time series (such as astrometry and radial velocities).
We define the full dataset as a concatenation of the astrometric and radial velocity datasets:
\begin{align}
  t & = (t_\mathrm{a}, t_\mathrm{rv}),\nonumber \\
  y & = (s, v),
\end{align}
where $t_\mathrm{a}$ and $t_\mathrm{rv}$ are the astrometric and radial velocities calendars respectively,
$s$ is the time series of astrometric abscissa, and $v$ is the radial velocity time series.
Vectors $t$ and $y$ both have length $n = n_\mathrm{a} + n_\mathrm{rv}$.
Since we assumed the noise affecting both time series to be independent,
the full covariance matrix is block-diagonal:
\begin{equation}
  C =   \begin{pmatrix}
    C_\mathrm{a} & 0             \\
    0            & C_\mathrm{rv}
  \end{pmatrix}.
\end{equation}
Considering the linear models $m_\mathrm{a}(\beta_\mathrm{a}) = \varphi_\mathrm{a}\beta_\mathrm{a}$ for the astrometric time series
and $m_\mathrm{rv}(\beta_\mathrm{rv}) = \varphi_\mathrm{rv}\beta_\mathrm{rv}$ for the radial velocity time series,
the corresponding full linear model is written as
\begin{equation}
  m(\beta) = \varphi \beta,
\end{equation}
with
\begin{align}
  \beta   & = (\beta_\mathrm{a}, \beta_\mathrm{rv}),\nonumber \\
  \varphi & =  \begin{pmatrix}
                 \varphi_\mathrm{a} & 0                   \\
                 0                  & \varphi_\mathrm{rv}
               \end{pmatrix},
\end{align}
and the corresponding $\chi^2$ is written as:
\begin{align}
  \chi^2 =\  & (y - \varphi\beta)\t C^{-1} (y - \varphi\beta)\nonumber                                                               \\
  =\         & (s - \varphi_\mathrm{a}\beta_\mathrm{a})\t C_\mathrm{a}^{-1} (s - \varphi_\mathrm{a}\beta_\mathrm{a})\nonumber        \\
             & + (v - \varphi_\mathrm{rv}\beta_\mathrm{rv})\t C_\mathrm{rv}^{-1} (v - \varphi_\mathrm{rv}\beta_\mathrm{rv})\nonumber \\
  =\         & \chi_\mathrm{a}^2 + \chi_\mathrm{rv}^2,
\end{align}
since $\varphi$ and $C$ are block-diagonal.

The base model $\H$ can thus be defined completely independently for the astrometry and the radial velocities.
As mentioned in Sect.~\ref{sec:periodef}, for the astrometry, the base model typically includes five linear parameters
(see Eq.~\ref{eq:basemodastro}).
In the radial velocity case, the base model typically includes instruments offsets and might additionally take into account drift terms
(typically linear, quadratic, or cubic trends) or activity indicators ($R'_{HK}$, FWHM, etc.).

The enlarged model $\K(\nu)$ includes four additional linear predictors for the astrometry
and two additional predictors for the radial velocities:
\begin{align}
  \varphi_{\K,\mathrm{a}}  & = (  \varphi_{\H, \mathrm{a}},
  \cos\theta\cos\nu t_\mathrm{a}, \sin\theta\cos\nu t_\mathrm{a},
  \cos\theta\sin\nu t_\mathrm{a}, \sin\theta\sin\nu t_\mathrm{a}),\nonumber                               \\
  \varphi_{\K,\mathrm{rv}} & = (\varphi_{\H, \mathrm{rv}}, \cos\nu t_\mathrm{rv}, \sin\nu t_\mathrm{rv}).
\end{align}
We thus have $d=6$ additional parameters in the model $\K(\nu)$, compared to $\H$.
The periodogram $z(\nu)$ is then computed according to the definition of Eq.~(\ref{eq:GLS}),
or one of the alternative definitions of Appendix~\ref{sec:fapcomputation}.
The FAP approximation is very similar to the case of astrometry alone (Sect.~\ref{sec:fap})
or radial velocities alone \citep[see][]{baluev_2008_assessing,delisle_2020_efficient}.
Furthermore, we have:
\begin{equation}
  \mathrm{FAP_{max}}(Z, \nu_\mathrm{max}) \approx 1 - \left(1-\mathrm{FAP_{single}}(Z)\right)\expo{-\tau(Z, \nu_\mathrm{max})},
\end{equation}
where analytical expressions for $\mathrm{FAP_{single}}$ and $\tau(Z, \nu_\mathrm{max})$
are given in Table~\ref{tab:power} ($d=6$).
The rescaled frequency bandwidth, $W$, is still defined as (see Eq.~(\ref{eq:defW}))
\begin{equation}
  W = \frac{\nu_\mathrm{max}}{2\pi} T_\mathrm{eff},
\end{equation}
but the effective time series length $T_\mathrm{eff}$
is now approximated as (see Appendix~\ref{sec:fapcomputation}):
\begin{align}
  \label{eq:Teffestimatecomb}
  T_\mathrm{eff} \approx\  & \frac{8\sqrt{\pi}}{15} \left(
  \bracktwolines{
    \sqrt{\mean\lambda_+} + \sqrt{\mean\lambda_-} + \sqrt{\mean\lambda_\mathrm{rv}}
  }{
    - \frac{
      \left(\sqrt{\mean\lambda_+\mean\lambda_-}
      + \sqrt{\mean\lambda_+\mean\lambda_\mathrm{rv}}
      + \sqrt{\mean\lambda_-\mean\lambda_\mathrm{rv}}\right)^2
    }{
      \left(\sqrt{\mean\lambda_+} + \sqrt{\mean\lambda_-}\right)
      \left(\sqrt{\mean\lambda_+} + \sqrt{\mean\lambda_\mathrm{rv}}\right)
      \left(\sqrt{\mean\lambda_-} + \sqrt{\mean\lambda_\mathrm{rv}}\right)
    }}\right).
\end{align}
The expressions of $\mean{\lambda}_+$, $\mean{\lambda}_-$, and $\mean{\lambda}_\mathrm{rv}$
are provided in Appendix~\ref{sec:fapcomputation} in the general case of correlated noise.
In the white noise case (diagonal covariance matrices $C_\mathrm{a} = \diag(\sigma_\mathrm{a}^2)$,
$C_\mathrm{rv} = \diag(\sigma_\mathrm{rv}^2)$), we find:
\begin{align}
  \label{eq:lambdawhite}
  \mean{\lambda}_+           & = \mean{\lambda}_- = \mean{t_\mathrm{a}^2}-\mean{t_\mathrm{a}}^2,\nonumber \\
  \mean{\lambda}_\mathrm{rv} & = \mean{t_\mathrm{rv}^2}-\mean{t_\mathrm{rv}}^2,
\end{align}
where $\mean{t_k}$ and $\mean{t_k^2}$ ($k = \mathrm{a}, \mathrm{rv}$)
are weighted means with weights $\sigma_{k,i}^2/\sum_j \sigma_{k,j}^2$.
The expression for the effective time series length is thus a combination of the
weighted standard deviations of the measurement times of both time series.

\section{Analytical determination of orbital parameters}
\label{sec:orbparam}

Once a significant peak has been identified in the periodogram,
we may attribute this peak to the presence of a companion.
The periodogram peak provides an estimate of the companion period
and the next step is usually to derive the other parameters.
This step is usually performed using numerical algorithms (local minimization methods, MCMC, genetic algorithms, etc.).
In order to improve the performances of such methods, a good initial guess of the parameters is very valuable.
We thus adapt the method we developed for the radial velocities \citep[see][]{delisle_2016_analytical}
to the case of an astrometric signal (Sect.~\ref{sec:guessastro})
and to the combination of astrometry and radial velocities (Sect.~\ref{sec:guesscomb}).
An analogous method was developed in the case of interferometric astrometry by \citet{konacki_2002_frequency}.
The principle of the method is to deduce the orbital parameters
from the amplitudes and phases of the signal at the period and first harmonics of the planet.
For instance, the amplitude of the $k$-th harmonics scales as $e^k$,
so the ratio between the amplitudes of the first harmonics and the fundamental period
provides insights into the eccentricity.

\subsection{Astrometric time series}
\label{sec:guessastro}

\begin{figure*}
  \centering
  \includegraphics[width=\linewidth]{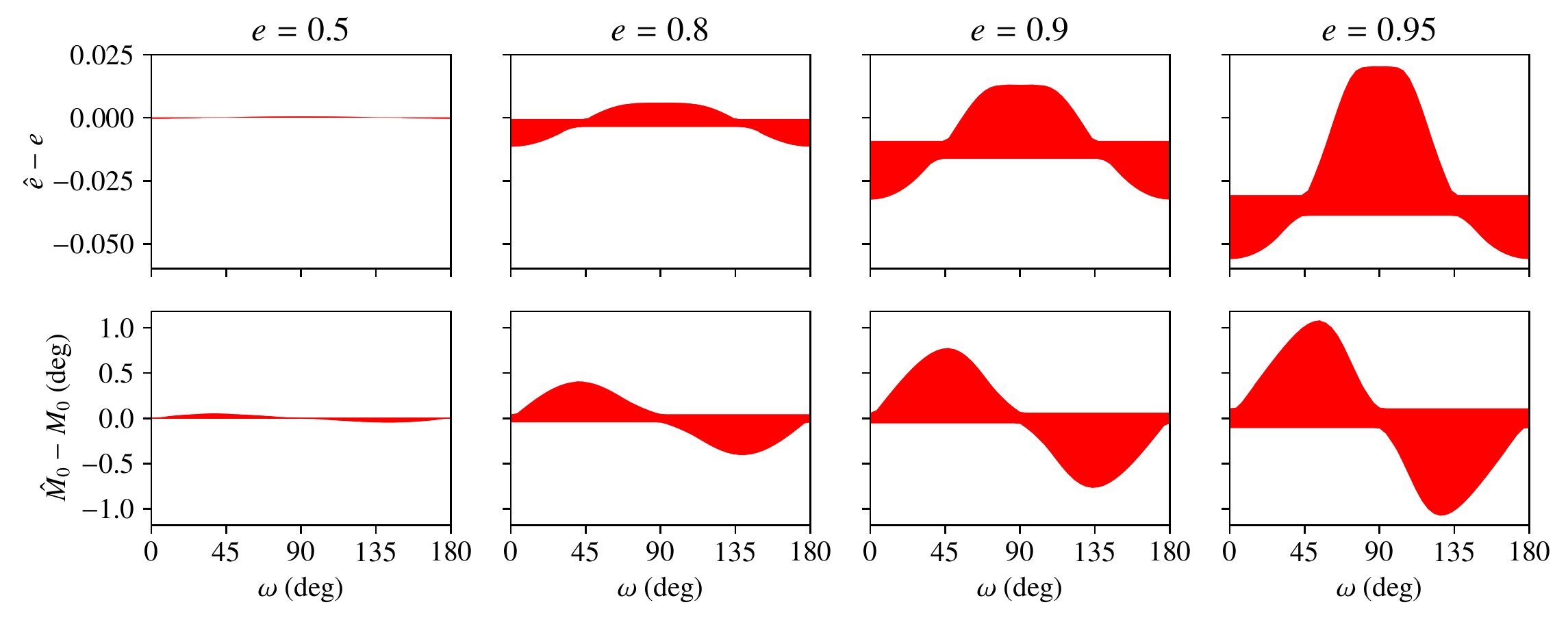}
  \caption{Accuracy of our analytical estimates of $e$ and $M_0$ (see Sect.~\ref{sec:guessastro}) assuming the Fourier decomposition of the signal to be perfectly known.
    We plot the range of deviations from the true eccentricity (\textit{top}) and mean anomaly (\textit{bottom}) as a function of the argument of periastron, $\omega$, and the true eccentricity, $e$.}
  \label{fig:perf_analytic}
\end{figure*}

We assume here that the orbital period of the companion has been correctly determined
with the periodogram approach.
We define a new linear model similar to the one used for the periodogram (Eq.~(\ref{eq:linmodnu}))
but including both the mean motion $n = 2\pi/P$ and the first harmonics $2 n$:
\begin{align}
  \varphi = ( & \varphi_\H,\nonumber                                   \\
              & \cos\theta\cos n t,\ \sin\theta\cos n t,\nonumber      \\
              & \cos\theta\sin n t,\ \sin\theta\sin n t,\nonumber      \\
              & \cos\theta\cos 2 n t,\ \sin\theta\cos 2 n t,\nonumber  \\
              & \cos\theta\sin 2 n t,\ \sin\theta\sin 2 n t),\nonumber \\
  \beta = (   & \beta_\H,\nonumber                                     \\
              & \beta_{\delta,1,c},\ \beta_{\alpha,1,c},\nonumber      \\
              & \beta_{\delta,1,s},\ \beta_{\alpha,1,s},\nonumber      \\
              & \beta_{\delta,2,c},\ \beta_{\alpha,2,c},\nonumber      \\
              & \beta_{\delta,2,s},\ \beta_{\alpha,2,s}).
\end{align}
The corresponding $\chi^2$ is written as
\begin{equation}
  \chi^2 = (s - \varphi\beta)\t C^{-1} (s - \varphi\beta),
\end{equation}
and is minimized for
\begin{equation}
  \beta = C_\beta \varphi\t C^{-1} s,
\end{equation}
where $C_\beta$ is the covariance matrix of the parameters
\begin{equation}
  \label{eq:Covbeta}
  C_\beta = \left(\varphi\t C^{-1} \varphi\right)^{-1}.
\end{equation}
By definition, we have (see Eq.~(\ref{eq:defdkak})):
\begin{align}
  \label{eq:da12}
  d_1 & \approx \frac{1}{2}\left(\beta_{\delta,1,c} - \i \beta_{\delta,1,s}\right),\nonumber \\
  a_1 & \approx \frac{1}{2}\left(\beta_{\alpha,1,c} - \i \beta_{\alpha,1,s}\right),\nonumber \\
  d_2 & \approx \frac{1}{2}\left(\beta_{\delta,2,c} - \i \beta_{\delta,2,s}\right),\nonumber \\
  a_2 & \approx \frac{1}{2}\left(\beta_{\alpha,2,c} - \i \beta_{\alpha,2,s}\right).
\end{align}
We then define (see Eqs.~(\ref{eq:dkak}),~(\ref{eq:zetak})):
\begin{align}
  \label{eq:ratio}
  \rho_\delta & = 2\frac{d_2}{d_1}
  = 2 \expo{\i M_0}\frac{\zeta_2 + \expo{-2\i \phi_\delta}\zeta_{-2}}{\zeta_1 + \expo{-2\i \phi_\delta}\zeta_{-1}}\nonumber            \\
              & = \expo{\i M_0}\left(e - \frac{e^3}{4} - \frac{e^3}{24} \expo{-2\i \phi_\delta} \right) + \O{e^5},\nonumber            \\
  \rho_\alpha & = 2\frac{a_2}{a_1}  = \expo{\i M_0}\left(e - \frac{e^3}{4} - \frac{e^3}{24} \expo{-2\i \phi_\alpha} \right) + \O{e^5},
\end{align}
and
\begin{align}
  \label{eq:ratio2}
  \eta_\delta & = 2\frac{d_2}{d_1^2} = \frac{e}{P_\delta} \expo{-\i \phi_\delta} + \O{e^3},\nonumber \\
  \eta_\alpha & = 2\frac{a_2}{a_1^2} = \frac{e}{P_\alpha} \expo{-\i \phi_\alpha} + \O{e^3}.
\end{align}
These expressions are very similar to the radial velocity case \citep[see][]{delisle_2016_analytical}.
We thus follow the same lines to deduce the planet's orbital parameters.
At first order in eccentricity, we find (see Eq.~(\ref{eq:ratio})):
\begin{equation}
  \rho_k \approx e \expo{\i M_0},
\end{equation}
so the eccentricity $e$ and mean anomaly $M_0$ could be directly deduced from the
amplitude and phase of $\rho_k$.
However, a better precision can be achieved by taking into account the terms of order 3 in eccentricity.
Following \citet{delisle_2016_analytical},
we first determine a crude approximation of $\phi_k$:
\begin{equation}
  \hat{\phi}_k \approx - \arg\left(\eta_k\right).
\end{equation}
Then, defining the coefficients $r_k$ as:
\begin{equation}
  r_k = \frac{1}{4}\left(1+\frac{\cos(2\hat{\phi}_k)}{6}\right),
\end{equation}
we obtain the same cubic equation for $e$ as in the radial velocity case
\citep[see Eq.~(\ref{eq:ratio}) and][]{delisle_2016_analytical}:
\begin{equation}
  |\rho_k| \approx e - r_k e^3,
\end{equation}
whose unique solution in the interval $[0, 1]$ is given by \citep[see][]{delisle_2016_analytical}:
\begin{equation}
  \hat{e}_k = \frac{1}{h_k} \cos\left(\frac{\pi + \arccos\left(3 h_k |\rho_k|\right)}{3}\right),
\end{equation}
with
\begin{equation}
  h_k = \frac{\sqrt{3r_k}}{2}.
\end{equation}
Then, the mean anomaly $M_0$ can be deduced using
(see Eq.~(\ref{eq:ratio})):
\begin{equation}
  \label{eq:M0da}
  \hat{M}_{0\,k} = \arg(\rho_k) - \arg\left(1-\frac{\hat{e}_k^2}{4}-\frac{\hat{e}_k^2}{24}\expo{-2\i\hat{\phi}_k}\right).
\end{equation}

With this method, we obtain two estimates for $e$ and $M_0$
-- one corresponding to the motion against the declination and one corresponding to the motion against the right ascension.
We then need to combine these estimates.
A simple average could be used, however, depending on the scanning law,
one of the estimates could be much more precise than the other.
We would thus rather perform a weighted average,
with weights computed by propagating the errorbars of the linear fit.
The details of this error propagation and weighting method are detailed in Appendix~\ref{sec:properror}.
In order to test the accuracy of our analytical estimate of $e$ and $M_0$,
we generated sets of orbital parameters and provided to our algorithm
the exact values of the Fourier decomposition
of the corresponding astrometric signal: $d_1$, $d_2$, $a_1$, $a_2$.
We then compared the eccentricity, $\hat{e}$, and mean anomaly, $\hat{M}_0$,
estimated from these Fourier coefficients using our algorithm,
with the true parameters $e$ and $M_0$.
We show in Fig.~\ref{fig:perf_analytic} the results of this comparison
as a function of the true eccentricity, $e$, and argument of periastron, $\omega$.
For each couple, $e$, $\omega$, we generated a grid of inclination, $i$, and longitude of the node, $\Omega$,
and plotted the maximum deviations over this grid in Fig.~\ref{fig:perf_analytic}.
We see that the analytical estimates of $e$ and $M_0$ remain very close
to the true values even at high eccentricity,
with a maximum deviation of about 0.05 for the eccentricity and 1~deg for the mean anomaly at $e=0.95$.
The method could be further refined by using a Newton-Raphson correcting algorithm,
as proposed in \citet{delisle_2016_analytical}.
However, in practice, the accuracy of the method is mainly limited
by the precision of the determination of the Fourier coefficients
$d_1$, $d_2$, $a_1$, $a_2$, which we neglected in this test.

Once $P$, $e$ and $M_0$ are known,
the orbit is then fully characterized by computing the Thiele-Innes coefficients,
which are linear parameters (see Eqs.~(\ref{eq:keplerianTI})-(\ref{eq:ThieleInnes})).
We first solve the Kepler equation (Eq.~(\ref{eq:Keplereq})) to compute the eccentric anomaly $E$.
Then we compute the vectors $x, y$ according to Eq.~(\ref{eq:xyfromeE}),
and solve the linear problem (see Eq.~(\ref{eq:keplerianTI})):
\begin{align}
  \varphi = ( & \varphi_\H,\nonumber                \\
              & x\cos\theta,\ x\sin\theta,\nonumber \\
              & y\cos\theta,\ y\sin\theta)
\end{align}
to obtain estimates for $A$, $B$, $F$, and $G$.

\subsection{Combining astrometric and radial velocity time series}
\label{sec:guesscomb}

The method described in Sect.~\ref{sec:guessastro}
to derive an estimate of the eccentricity, $e$, and of the mean anomaly at reference epoch, $M_0$,
can be extended to the case of a combination of astrometric and radial velocity time series in a straightforward way.
We still assume that the orbital period of the companion
has been correctly determined with the periodogram approach
and we define a linear model including the fundamental frequency (mean motion $n=2\pi/P$) and the first harmonics ($2n$),
for both time series:
\begin{align}
  \varphi_\mathrm{a} = (  & \varphi_{\H,\mathrm{a}},\nonumber                                            \\
                          & \cos\theta\cos n t_\mathrm{a},\ \sin\theta\cos n t_\mathrm{a},\nonumber      \\
                          & \cos\theta\sin n t_\mathrm{a},\ \sin\theta\sin n t_\mathrm{a},\nonumber      \\
                          & \cos\theta\cos 2 n t_\mathrm{a},\ \sin\theta\cos 2 n t_\mathrm{a},\nonumber  \\
                          & \cos\theta\sin 2 n t_\mathrm{a},\ \sin\theta\sin 2 n t_\mathrm{a}),\nonumber \\
  \beta_\mathrm{a} = (    & \beta_{\H,\mathrm{a}},\nonumber                                              \\
                          & \beta_{\delta,1,c},\ \beta_{\alpha,1,c},\nonumber                            \\
                          & \beta_{\delta,1,s},\ \beta_{\alpha,1,s},\nonumber                            \\
                          & \beta_{\delta,2,c},\ \beta_{\alpha,2,c},\nonumber                            \\
                          & \beta_{\delta,2,s},\ \beta_{\alpha,2,s}),\nonumber                           \\
  \varphi_\mathrm{rv} = ( & \varphi_{\H,\mathrm{rv}},\nonumber                                           \\
                          & \cos n t_\mathrm{rv},\ \sin n t_\mathrm{rv},\nonumber                        \\
                          & \cos 2 n t_\mathrm{rv},\ \sin 2 n t_\mathrm{rv}),\nonumber                   \\
  \beta_\mathrm{rv} = (   & \beta_{\H,\mathrm{rv}},\nonumber                                             \\
                          & \beta_{\mathrm{rv},1,c},\ \beta_{\mathrm{rv},1,s},\nonumber                  \\
                          & \beta_{\mathrm{rv},2,c},\ \beta_{\mathrm{rv},2,s}).
\end{align}
We obtain the best-fitting parameters $\beta_\mathrm{a}$, $\beta_\mathrm{rv}$,
as well as the covariance matrices $C_{\beta_\mathrm{a}}$, $C_{\beta_\mathrm{rv}}$
by minimizing the corresponding $\chi^2$.
We derive the values of $\rho_\delta$, $\rho_\alpha$, $\eta_\delta$, $\eta_\alpha$ (see Eqs.~(\ref{eq:ratio})-(\ref{eq:da12}))
from the astrometric parameters $\beta_\mathrm{a}$.
We also derive \citep[see][]{delisle_2016_analytical}:
\begin{align}
  \rho_\mathrm{rv} & = \frac{V_2}{V_1},\nonumber \\
  \eta_\mathrm{rv} & = \frac{V_2}{V_1^2},
\end{align}
with
\begin{align}
  V_1 & = \frac{1}{2}(\beta_{\mathrm{rv},1,c} - \i \beta_{\mathrm{rv},1,s}),\nonumber \\
  V_2 & = \frac{1}{2}(\beta_{\mathrm{rv},2,c} - \i \beta_{\mathrm{rv},2,s}).
\end{align}
We then use the method described in \citet{delisle_2016_analytical} and Sect.~\ref{sec:guessastro}
to obtain the estimates $\hat{e}_k$ and $\hat{M}_{0\,k}$ from the values of $\rho_k$ and $\eta_k$ ($k=\delta,\alpha,\mathrm{rv}$).
These three estimates are finally combined to obtain $\hat{e}$ and $\hat{M}_0$,
using the error propagation and weighting method detailed in Appendix~\ref{sec:properror}.

Following the same line as in Sect.~\ref{sec:guessastro},
we performed a linear fit of the astrometric time series to obtain
estimates of the Thiele-Innes coefficients ($A$, $B$, $F$, and $G$).
Similarly, we performed a linear fit of the radial velocity time series to obtain
estimates of $K_c = K\cos\omega$ and $K_s = K\sin\omega$.
These estimates should again be combined to obtain a set of four parameters describing the orbit,
in addition to the three parameters already determined ($P$, $e$, $M_0$).

The astrometry is sensitive to all the orbital parameters but has a degeneracy
between two symmetric solutions
for the argument of periastron $\omega$ and the longitude of the node $\Omega$
($\omega'=\omega+\pi$, $\Omega'=\Omega+\pi$).
On the contrary, the radial velocities are not sensitive to the inclination, $i$, and the longitude of the node, $\Omega$,
but allow us to break the degeneracy by providing an estimate of the argument of periastron, $\omega$.
To combine the parameters obtained from astrometry and radial velocities,
we introduce the following parameters:
\begin{align}
  U & = (a_{s,\mathrm{AU}} \sin i)^2 \cos 2\omega,\nonumber \\
  V & = (a_{s,\mathrm{AU}} \sin i)^2 \cos 2\omega.
\end{align}
These parameters can be derived both from the astrometric parameters,
\begin{align}
  \hat{U}_\mathrm{a} & = \frac{(\hat{A}^2+\hat{B}^2)-(\hat{F}^2+\hat{G}^2)}{\hat{\varpi}^2},\nonumber \\
  \hat{V}_\mathrm{a} & = -2\frac{\hat{A}\hat{F}+\hat{B}\hat{G}}{\hat{\varpi}^2},
\end{align}
and the radial velocity parameters,
\begin{align}
  \hat{U}_\mathrm{rv} & = \frac{1-\hat{e}^2}{\hat{n}^2} \left(\hat{K}_c^2 - \hat{K}_s^2\right),\nonumber \\
  \hat{V}_\mathrm{rv} & = 2\frac{1-\hat{e}^2}{\hat{n}^2} \hat{K}_c \hat{K}_s.
\end{align}
We followed the same approach as in Appendix~\ref{sec:properror} to propagate the covariance matrices of the
linear fits and performed a weighted average of the two estimates of $U$ and $V$.
From the weighted estimates $\hat{U}$, $\hat{V}$, it is straightforward to derive estimates for $a_s\sin i$ and $\omega$.
For the argument of periastron, $\omega$, two solutions are actually possible since $U$ and $V$ depend on $2\omega$,
but we chose the one that is closest to the radial velocity estimate (obtained from $\hat{K}_c$, $\hat{K}_s$).
The two remaining parameters only depend on the astrometry and can be chosen arbitrarily.
For instance, $\Omega$ can be estimated with
\begin{equation}
  \hat{\Omega} = \arctan\left(\hat{B}\cos\omega-\hat{G}\sin\omega, \hat{A}\cos\omega-\hat{F}\sin\omega\right),
\end{equation}
and $\cos i$ can be estimated using \citep[e.g.,][]{popovic_1995_approach}:
\begin{equation}
  \cos\hat{i} = \frac{m}{k+j},
\end{equation}
with
\begin{align}
  m & = \hat{A}\hat{G} - \hat{B}\hat{F}
  \approx a_s^2 \cos i,\nonumber                                              \\
  k & = \frac{1}{2}\left(\hat{A}^2 + \hat{B}^2 + \hat{F}^2 + \hat{G}^2\right)
  \approx a_s^2 \frac{1 + \cos^2 i}{2},\nonumber                              \\
  j & = \sqrt{k^2-m^2} \approx a_s^2 \frac{\sin^2 i}{2}.
\end{align}

\section{Applications}
\label{sec:applications}

We go on to illustrate our methods with a reanalysis of the astrometric and radial velocity time series
of three stars known to host a companion:
\object{HD~223636} (HIP~117622), \object{HD~17289} (HIP~12726), and \object{HD~3277} (HIP~2790).
We used Hipparcos data \citep[with the reduction of][]{vanleeuwen_2007_hipparcos} for the astrometry and
CORALIE data for the radial velocities.
We binned Hipparcos data over one-day intervals and excluded the bins with a resulting uncertainty larger than 2.5~mas.

An upgrade of CORALIE was performed in 2007 to improve the overall transmission of the instrument.
This upgrade affected the instrumental zero point and the overall instrument stability \citep{segransan_2010_coralie}.
We thus actually considered the data taken before and after the upgrade as coming from two
different instruments (COR98 before and COR07 after).
We modeled the noise as the quadratic sum of the individual measurements errors,
as provided by the instruments reduction pipelines,
and an additional instrumental jitter: $\tilde\sigma_i = \sqrt{\sigma_i^2 + \sigma_\mathrm{inst.}^2}$.
Such a modeling process allows us to account for both stellar and instrumental noises
that are not already modeled by the reduction pipelines.
We thus defined three jitter terms: $\sigma_\mathrm{HIP}$ for Hipparcos data
and then $\sigma_\mathrm{COR98}$ and $\sigma_\mathrm{COR07}$ for CORALIE data.
We initially set the default values
$\sigma_\mathrm{COR98} = 5\ \mathrm{m/s}$
and $\sigma_\mathrm{COR07} = 8\ \mathrm{m/s}$,
which correspond to the observed scatter
of a set of bright and stable solar type stars monitored
on a regular basis with CORALIE since 1998.
For Hipparcos, we used a default value of $\sigma_\mathrm{HIP} = 0\ \mathrm{mas}$.
We let the three jitter values vary at the end of the analysis to adapt to the specificity of each dataset.

We applied a similar procedure for each target, using our open source python package kepmodel\footnote{\kepmodelURL}.
We analyzed the astrometric and radial velocity time series both independently and jointly.
We computed the periodogram and associated FAP according to \citet{delisle_2020_efficient} (radial velocities)
and Sect.~\ref{sec:periofap} (astrometry and joint analysis).
All the periodograms were computed using the normalized power of Eq.~(\ref{eq:GLS})
and by sampling the angular frequency linearly
in the range $[2\pi/50000, 2\pi/0.9]$ (rad/d) with 50000 values,
such that the oversampling factor is always above 10.
We then extracted from the periodogram the period corresponding to the highest peak and applied our
analytical methods to estimate the other orbital parameters, using
\citet{delisle_2016_analytical} (radial velocities)
and Sect.~\ref{sec:orbparam} (astrometry and joint analysis).
We then adjusted the values of all the parameters by performing a local maximization of the likelihood $\mathcal{L}$ using the L-BFGS-B algorithm provided by the scipy python package.
This algorithm makes use of the partial derivatives of the likelihood with respect to the parameters,
which we computed analytically.
We estimated errorbars on the parameters by extracting the diagonal of the inverse of the Hessian matrix of the likelihood.
We then repeated the fitting procedure but additionally adjusting the values of the instrumental jitter terms.
Finally, we computed the periodograms and FAP of the residuals of the joint model,
considering the astrometric and radial velocity time series independently or jointly.

\subsection{HD~223636}
\label{sec:hd223636}

\begin{figure}
  \centering
  \includegraphics[width=\linewidth]{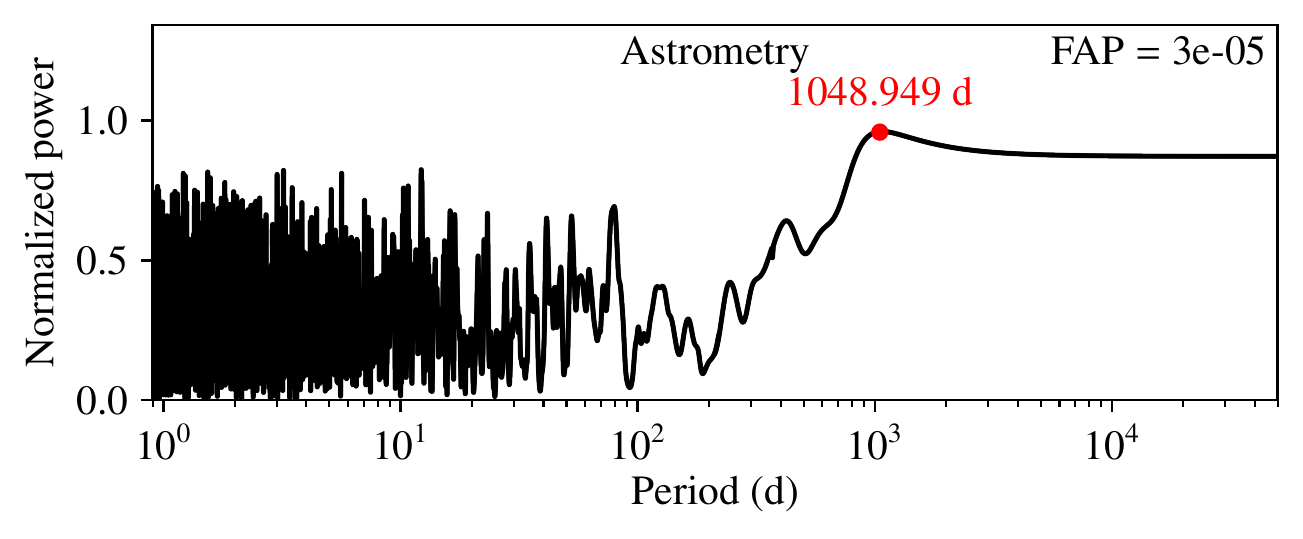}
  \caption{Periodogram of the Hipparcos astrometric time series of HD~223636.}
  \label{fig:hd223636_perio_0}
\end{figure}

\begin{table}
  \caption{Fit parameters for HD~223636.}
  \begin{tabular}{lccc}
    \hline
    \hline
                                & \multicolumn{3}{c}{Astrometry}                                                                     \\
                                & guess                          & fit                             & fit incl. jitter                \\
    \hline
    $\pi$ (mas)                 & $       19.45$                 & $       19.66 \pm         0.39$ & $       19.70 \pm         0.51$ \\
    $\Delta \delta$ (mas)       & $       -33.8$                 & $       -32.3 \pm          1.6$ & $       -31.8 \pm          2.1$ \\
    $\Delta \alpha^*$ (mas)     & $        77.2$                 & $        78.1 \pm          1.0$ & $        78.5 \pm          1.3$ \\
    $\mu_\delta$ (mas/yr)       & $      -75.56$                 & $      -75.35 \pm         0.42$ & $      -75.35 \pm         0.53$ \\
    $\mu_\alpha$ (mas/yr)       & $      197.58$                 & $      197.96 \pm         0.84$ & $       198.2 \pm          1.0$ \\
    $\sigma_\mathrm{HIP}$ (mas) & --                             & --                              & $        0.64 \pm         0.21$ \\
    \hline
    $P$ (d)                     & $        1049$                 & $        1011 \pm           32$ & $        1004 \pm           37$ \\
    $M_0+\omega$ (deg)          & $       242.2$                 & $       243.4 \pm          3.4$ & $       243.0 \pm          4.2$ \\
    $a_\mathrm{s}$ (mas)        & $       12.37$                 & $       12.27 \pm         0.43$ & $       12.48 \pm         0.59$ \\
    $e$                         & $       0.272$                 & $       0.257 \pm        0.059$ & $       0.261 \pm        0.086$ \\
    $\omega$ (deg)              & $         125$                 & $         100 \pm           24$ & $          92 \pm           29$ \\
    $i$ (deg)                   & $       118.2$                 & $       118.0 \pm          2.7$ & $       117.1 \pm          3.9$ \\
    $\Omega$ (deg)              & $        38.0$                 & $        37.7 \pm          2.7$ & $        37.3 \pm          3.4$ \\
    \hline
    $\log\mathcal{L}$           & -40.00                         & -39.41                          & -35.81                          \\
    \hline
  \end{tabular}
  \tablefoot{The reference epoch is 2\,448\,500~BJD.}
  \label{tab:hd223636}
\end{table}

\begin{figure}
  \centering
  \includegraphics[width=0.7\linewidth]{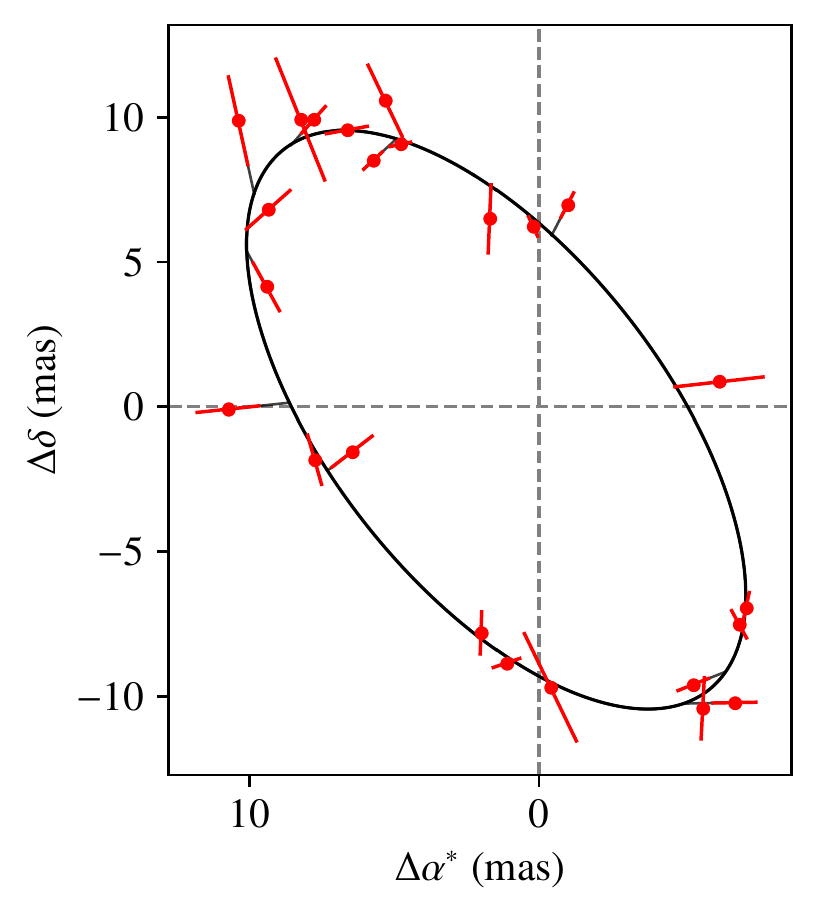}
  \caption{Astrometric orbit of HD~223636.}
  \label{fig:hd223636_astro_orbit}
\end{figure}

\begin{figure}
  \centering
  \includegraphics[width=\linewidth]{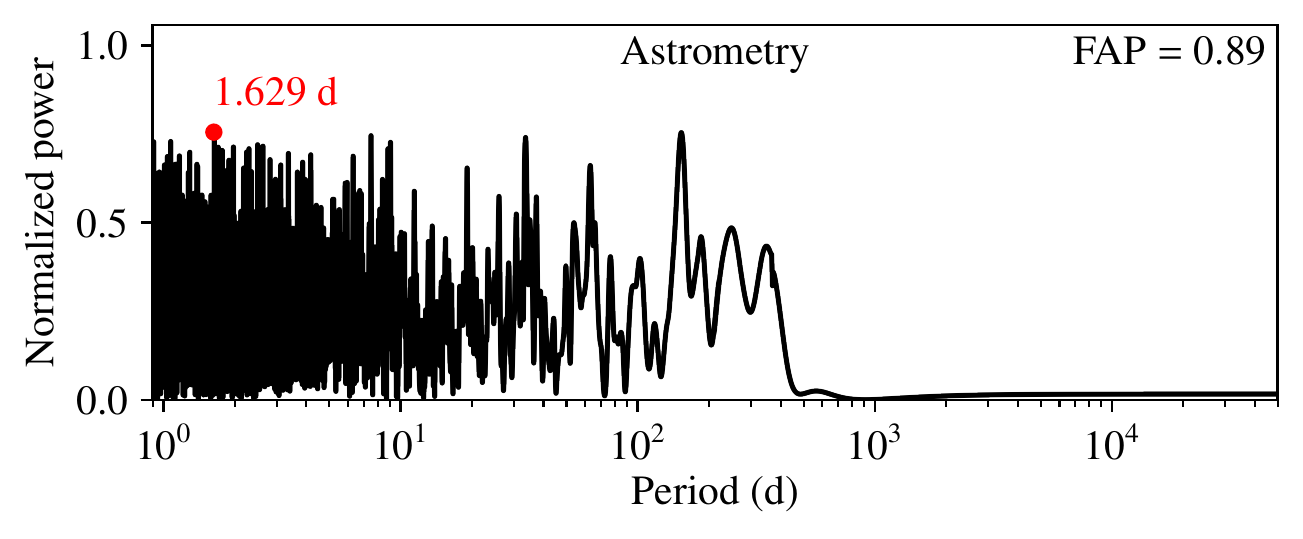}
  \caption{Periodogram of Hipparcos astrometric residuals of HD~223636
    after fitting for its companion.}
  \label{fig:hd223636_perio_1}
\end{figure}

We first considered HD~223636 (HIP~117622),
which was found to host a companion at about 1008~d by \citet{goldin_2007_astrometric},
based on an analysis of the Hipparcos time series.
No radial velocities are available for this target and we thus only reanalyzed the Hipparcos time series.
The periodogram of the astrometric data is shown in Fig.~\ref{fig:hd223636_perio_0},
where we clearly see a very significant peak ($\mathrm{FAP}=3\times10^{-5}$) around 1050~d.

We show in Table~\ref{tab:hd223636} the parameters obtained using our analytical estimation method
(initial guess),
after a local maximization of the likelihood and after a second fit, including the jitter term.
We observe that the initial guess of the parameters is very close to the maximum likelihood estimate.
When adjusting the jitter, we obtain a value of about 0.64~mas,
while the average Hipparcos errorbar (over one day bins) is about 0.93~mas.
Accounting for this jitter significantly increases the uncertainties on all the parameters,
and allows to have a better assessment of the actual precision on these parameters.
We note that the final uncertainty on the argument of periastron, $\omega$, is large (29~deg),
while the other angles have errorbars below 5~deg.
The set of parameters is chosen as to avoid strong correlations between these angles.
In particular, the phase of the planet on its orbit is modeled using
the mean argument at the reference epoch $M_0+\omega$
instead of the mean anomaly $M_0$ (which is anti-correlated with $\omega$)
or the mean longitude $\lambda_0=M_0+\omega+\Omega$ (which is correlated with $\Omega$).
The astrometric orbit corresponding to the maximum likelihood parameters (including jitter),
is shown in Fig.~\ref{fig:hd223636_astro_orbit} superimposed with Hipparcos measurements.
The periodogram of the residuals is shown in Fig.~\ref{fig:hd223636_perio_1},
where no significant peak is found ($\mathrm{FAP}=0.89$).

\subsection{HD~17289}
\label{sec:hd17289}

\begin{figure}
  \centering
  \includegraphics[width=\linewidth]{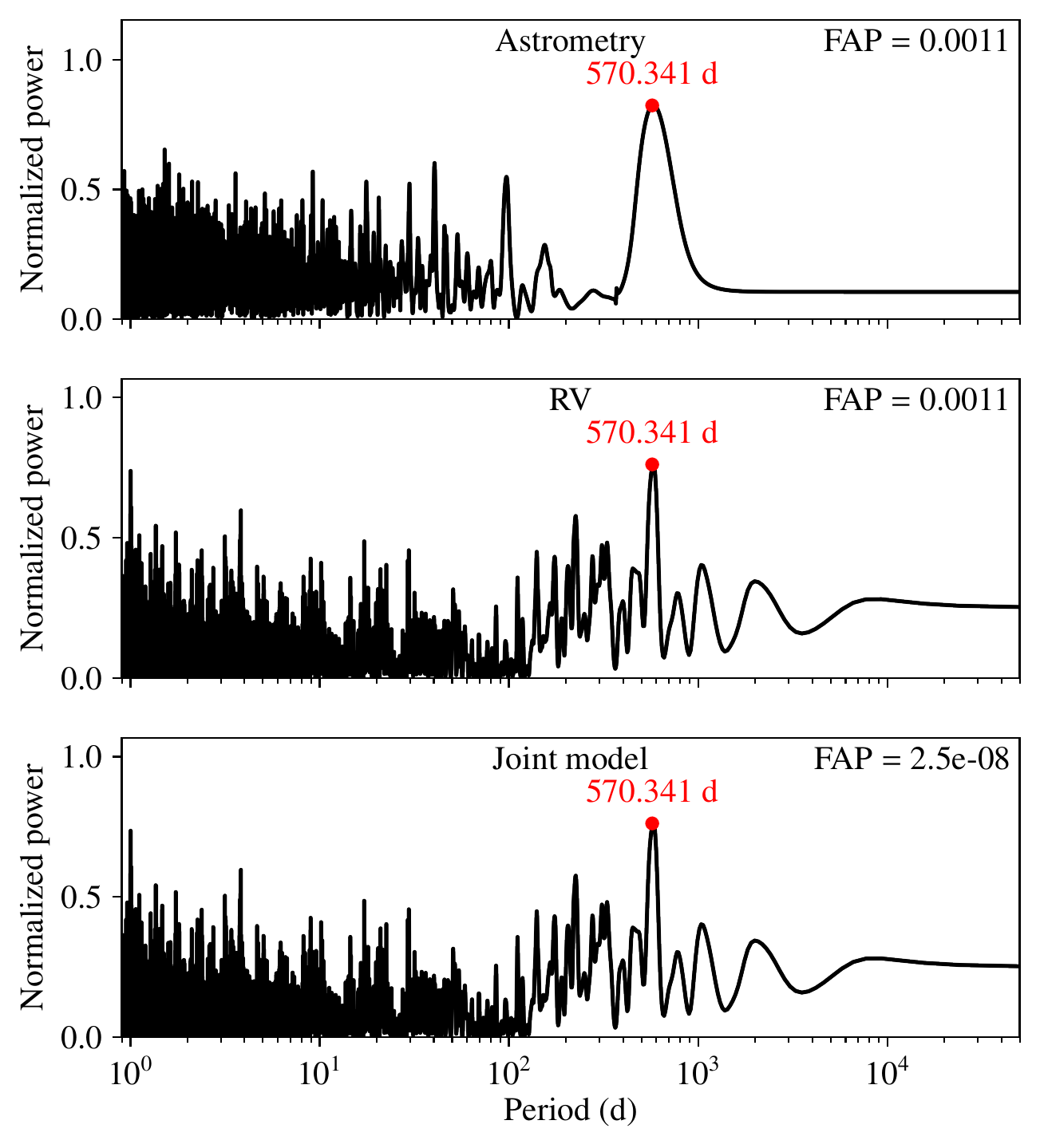}
  \caption{Periodograms of the Hipparcos time series (astrometry, \textit{top}),
    CORALIE time series (RV, \textit{middle}),
    and joint periodogram (\textit{bottom}) of HD~17289.}
  \label{fig:hd17289_perio_0}
\end{figure}

\begin{table*}
  \caption{Fit parameters of HD~17289.}
  \resizebox{\textwidth}{!}{\setlength{\tabcolsep}{1mm}
    \begin{tabular}{lccccccccc}
      \hline
      \hline
                                    & \multicolumn{3}{c}{Astrometry} & \multicolumn{3}{c}{RV}          & \multicolumn{3}{c}{Joint model}                                                                                                                                                                                     \\
                                    & guess                          & fit                             & fit incl. jitter                & guess          & fit                                  & fit incl. jitter                     & guess          & fit                             & fit incl. jitter                \\
      \hline
      $\pi$ (mas)                   & $       20.73$                 & $       20.41 \pm         0.34$ & $       20.34 \pm         0.56$ & --             & --                                   & --                                   & $       20.39$ & $       20.49 \pm         0.33$ & $       20.39 \pm         0.58$ \\
      $\Delta \delta$ (mas)         & $      -175.0$                 & $      -170.6 \pm          2.3$ & $      -170.4 \pm          3.4$ & --             & --                                   & --                                   & $      -167.3$ & $      -170.9 \pm          1.9$ & $      -170.3 \pm          3.1$ \\
      $\Delta \alpha^*$ (mas)       & $      -129.3$                 & $      -128.9 \pm          2.3$ & $      -128.6 \pm          3.8$ & --             & --                                   & --                                   & $      -118.4$ & $      -125.5 \pm          1.7$ & $      -125.1 \pm          2.7$ \\
      $\mu_\delta$ (mas/yr)         & $      -39.27$                 & $      -39.42 \pm         0.41$ & $      -39.42 \pm         0.65$ & --             & --                                   & --                                   & $      -37.74$ & $      -39.35 \pm         0.40$ & $      -39.29 \pm         0.68$ \\
      $\mu_\alpha$ (mas/yr)         & $      -26.37$                 & $      -26.22 \pm         0.38$ & $      -26.16 \pm         0.59$ & --             & --                                   & --                                   & $      -26.40$ & $      -26.26 \pm         0.36$ & $      -26.13 \pm         0.59$ \\
      $\sigma_\mathrm{HIP}$ (mas)   & --                             & --                              & $        1.43 \pm         0.34$ & --             & --                                   & --                                   & --             & --                              & $        1.57 \pm         0.34$ \\
      \hline
      $\gamma_\mathrm{COR98}$ (m/s) & --                             & --                              & --                              & $     34931.9$ & $     34942.1 \pm          2.3$      & $     34942.1 \pm          5.0$      & $     34931.5$ & $     34942.4 \pm          2.3$ & $     34942.3 \pm          5.0$ \\
      $\gamma_\mathrm{COR07}$ (m/s) & --                             & --                              & --                              & $     35187.0$ & $     34957.6 \pm          6.6$      & $     34954.8 \pm          9.1$      & $     35186.8$ & $     34956.4 \pm          6.5$ & $     34954.2 \pm          9.1$ \\
      $\sigma_\mathrm{COR98}$ (m/s) & --                             & --                              & --                              & --             & --                                   & $        17.9 \pm          3.5$      & --             & --                              & $        17.9 \pm          3.5$ \\
      $\sigma_\mathrm{COR07}$ (m/s) & --                             & --                              & --                              & --             & --                                   & $         0.0 \pm          6.6$      & --             & --                              & $         0.0 \pm          6.5$ \\
      \hline
      $P$ (d)                       & $       570.3$                 & $       567.2 \pm          4.7$ & $       570.5 \pm          7.0$ & $      570.34$ & $      562.03 \pm         0.19$      & $      562.00 \pm         0.27$      & $      570.34$ & $      561.99 \pm         0.18$ & $      561.98 \pm         0.27$ \\
      $M_0+\omega$ (deg)            & $         135$                 & $         141 \pm           16$ & $         129 \pm           29$ & $      185.53$ & $      162.40 \pm         0.54$      & $      162.71 \pm         0.80$      & $      185.59$ & $      162.28 \pm         0.53$ & $      162.65 \pm         0.80$ \\
      $a_\mathrm{s}$ (AU)           & $       0.662$                 & $       0.665 \pm        0.064$ & $        0.66 \pm         0.10$ & $     0.06635$ & $     0.06181 \pm 1.7\times 10^{-4}$ & $     0.06176 \pm 3.4\times 10^{-4}$ & $       0.157$ & $       0.557 \pm        0.019$ & $       0.576 \pm        0.034$ \\
      $e$                           & $       0.596$                 & $       0.656 \pm        0.061$ & $        0.65 \pm         0.10$ & $      0.5641$ & $      0.5326 \pm       0.0020$      & $      0.5311 \pm       0.0039$      & $      0.5637$ & $      0.5330 \pm       0.0020$ & $      0.5314 \pm       0.0039$ \\
      $\omega$ (deg)                & $          71$                 & $          55 \pm           15$ & $          46 \pm           29$ & $       54.26$ & $       52.24 \pm         0.30$      & $       53.04 \pm         0.40$      & $       54.40$ & $       52.27 \pm         0.30$ & $       53.06 \pm         0.40$ \\
      $i$ (deg)                     & $       149.0$                 & $       151.3 \pm          7.6$ & $         154 \pm           13$ & --             & --                                   & --                                   & $      154.98$ & $      173.63 \pm         0.22$ & $      173.84 \pm         0.36$ \\
      $\Omega$ (deg)                & $         348$                 & $         353 \pm           14$ & $         346 \pm           27$ & --             & --                                   & --                                   & $        48.0$ & $         4.3 \pm          1.9$ & $         5.2 \pm          3.1$ \\
      \hline
      $\log\mathcal{L}$             & -81.14                         & -72.66                          & -65.20                          & -1496.57       & -141.12                              & -117.52                              & -1858.53       & -218.30                         & -184.31                         \\
      \hline
    \end{tabular}
  }
  \tablefoot{The reference epoch is 2\,450\,000~BJD.}
  \label{tab:hd17289}
\end{table*}

\begin{figure}
  \centering
  \includegraphics[width=0.75\linewidth]{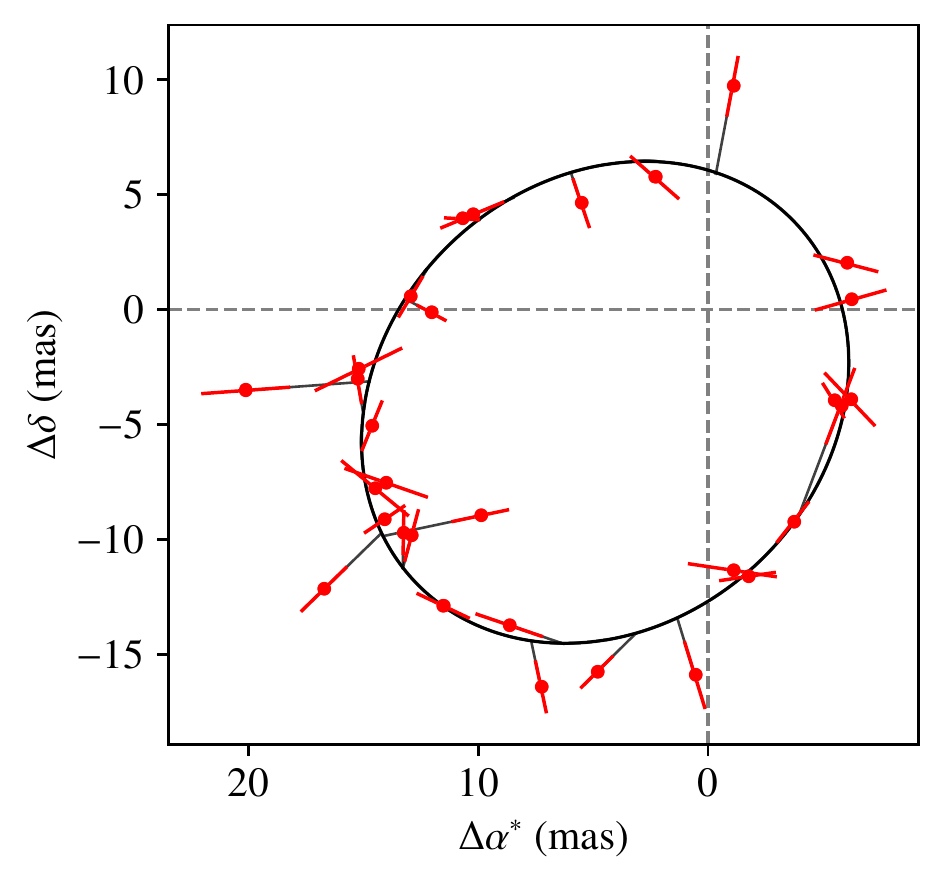}
  \caption{Astrometric orbit of HD~17289.}
  \label{fig:hd17289_astro_orbit}
\end{figure}

\begin{figure}
  \centering
  \includegraphics[width=\linewidth]{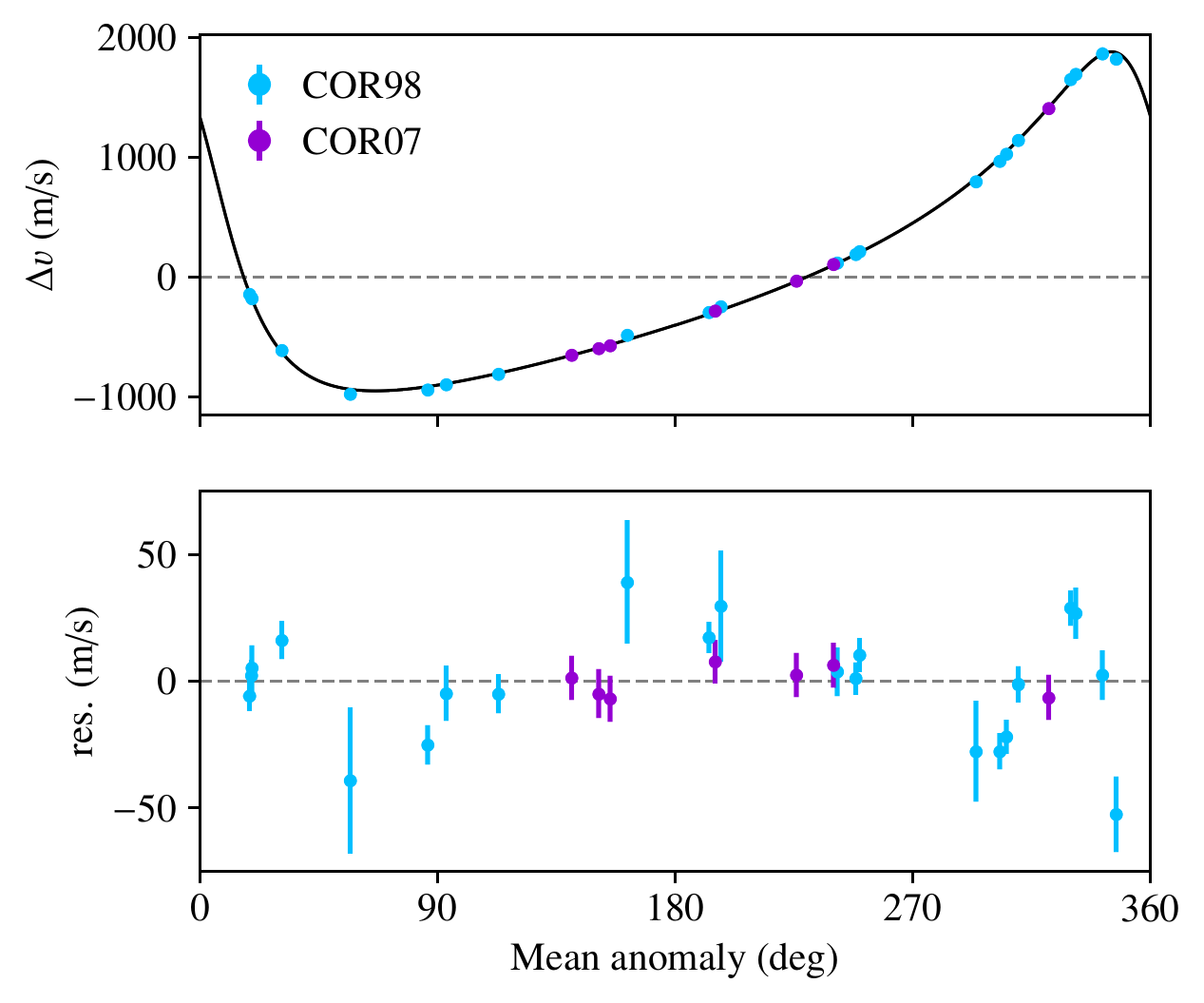}
  \caption{Phase-folded RV time series of HD~17289.}
  \label{fig:hd17289_rv_orbit}
\end{figure}

\begin{figure}
  \centering
  \includegraphics[width=\linewidth]{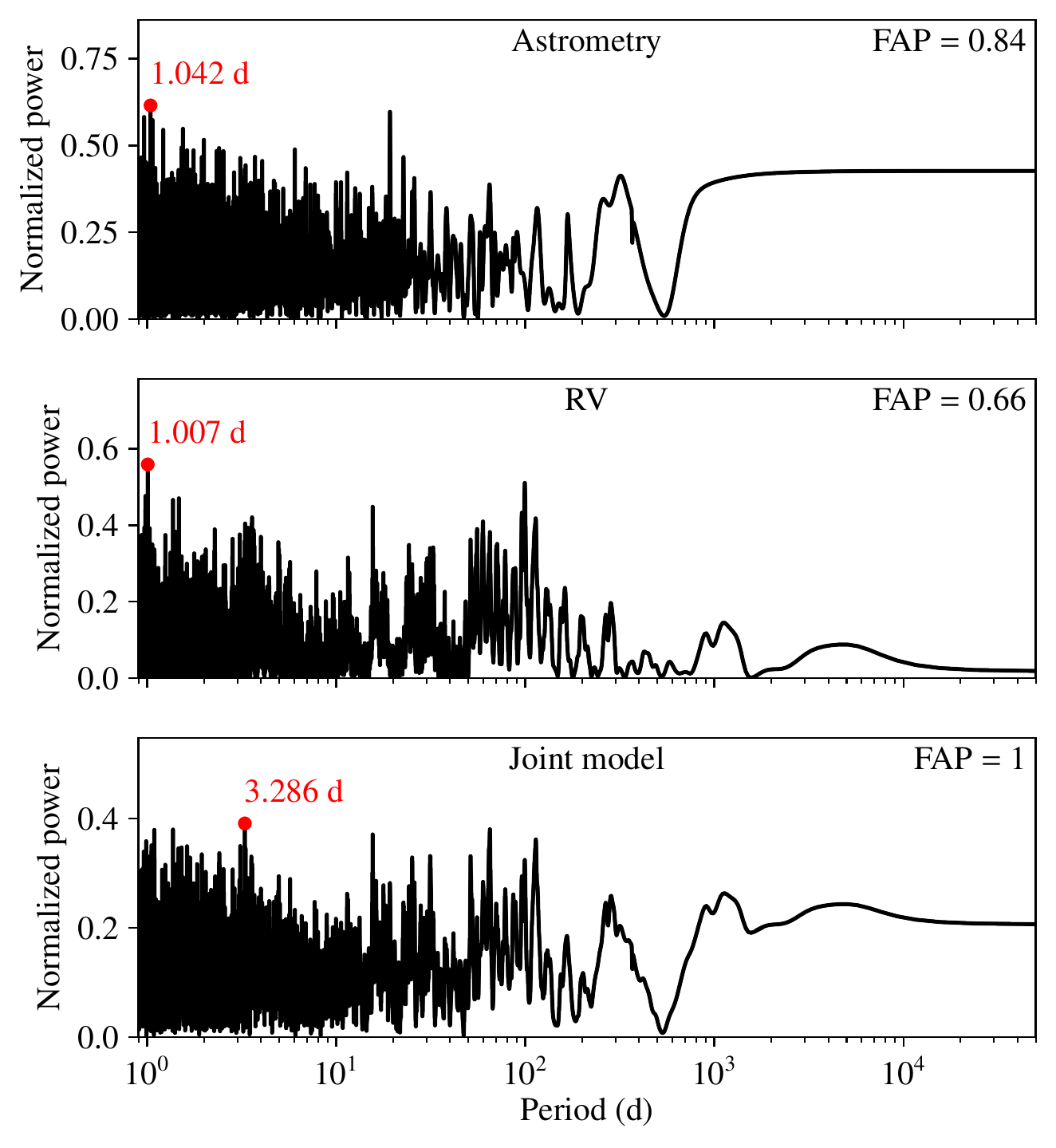}
  \caption{Periodograms of the residuals of
    the Hipparcos time series (astrometry, \textit{top}),
    CORALIE time series (RV, \textit{middle}),
    and joint periodogram (\textit{bottom}) of HD~17289,
    after fitting for its companion.}
  \label{fig:hd17289_perio_1}
\end{figure}

The second target, HD~17289 (HIP~12726), was also found to host a companion
by \citet{goldin_2007_astrometric} using Hipparcos astrometry alone.
The companion was later confirmed and better characterized
by \citet{sahlmann_2011_search} using both Hipparcos astrometry and CORALIE radial velocities.
This system is actually a binary star, with a mass ratio of about 2 between the primary and the companion,
leading to a contrast of about $1:180$ in the visible,
which slightly affects the radial-velocity measurements.
This issue is discussed in details in \citet{sahlmann_2011_search} and is beyond the scope of our study.
Here, we use the same Hipparcos and CORALIE data as \citet{sahlmann_2011_search}.
The CORALIE data consist of 29 measurements (22 COR98 and 7 COR07).

We show in Fig.~\ref{fig:hd17289_perio_0} the periodogram of the astrometric data alone,
the radial velocities alone, as well as the joint periodogram.
In the three cases, a significant peak is detected around 570~d.
Since the time span of CORALIE measurements (11.3~yr) is much longer than
the time span of Hipparcos measurements (3.2~yr),
the radial velocity peak is much narrower.
The joint periodogram shows a very similar shape as the radial velocity periodogram,
but provides stronger evidence for the existence of a companion
($\mathrm{FAP} = 3\times 10^{-8}$ and $10^{-3}$, respectively).

Table~\ref{tab:hd17289} provides the parameters obtained at the different steps of our procedure
using astrometry alone, radial velocities alone, and both datasets together.
In all three cases, our analytical method provides a good initial guess of the parameters.
When using the radial velocities alone, the provided value for the semi-major axis, $a_s$,
is actually the minimum semi-major axis ($a_s \sin i$), which is equivalent to assuming an inclination of 90~deg.
Since the actual inclination is about 174~deg ($\sin i \approx 0.1$),
this explains the factor of 10 difference with regard to the astrometric or joint value.
The other parameters show coherent values between astrometry and radial velocities.
When using astrometric data alone, the parameters $\omega$ and $\Omega$
present two degenerated solutions $(\omega,\ \Omega)$ and $(\omega + \pi,\ \Omega + \pi)$.
In order to ease the comparisons,
we chose the solution that is the closest to the value of $\omega$ obtained with radial velocities.
However, it should be kept in mind that there is no way to break this degeneracy using astrometry alone.
We observe that while the inclination, $i$, and longitude of the node $\Omega$ are not constrained by radial velocities,
we obtain a much higher precision on these parameters with the joint fit than using astrometry alone.
This is because of the correlations between these two parameters and the other orbital parameters,
which are much better constrained by the radial velocities than the astrometry.

The astrometric orbit corresponding to the maximum likelihood parameters (including jitter),
superimposed with Hipparcos measurements,
is shown in Fig.~\ref{fig:hd17289_astro_orbit},
while the phase-folded radial velocity model, superimposed with CORALIE measurements,
is shown in Fig.~\ref{fig:hd17289_rv_orbit}.
Finally, the periodograms of the residuals are shown in Fig.~\ref{fig:hd17289_perio_1},
where no significant peak is found ($\mathrm{FAP}>0.66$).

\subsection{HD~3277}
\label{sec:hd3277}

\begin{figure}
  \centering
  \includegraphics[width=\linewidth]{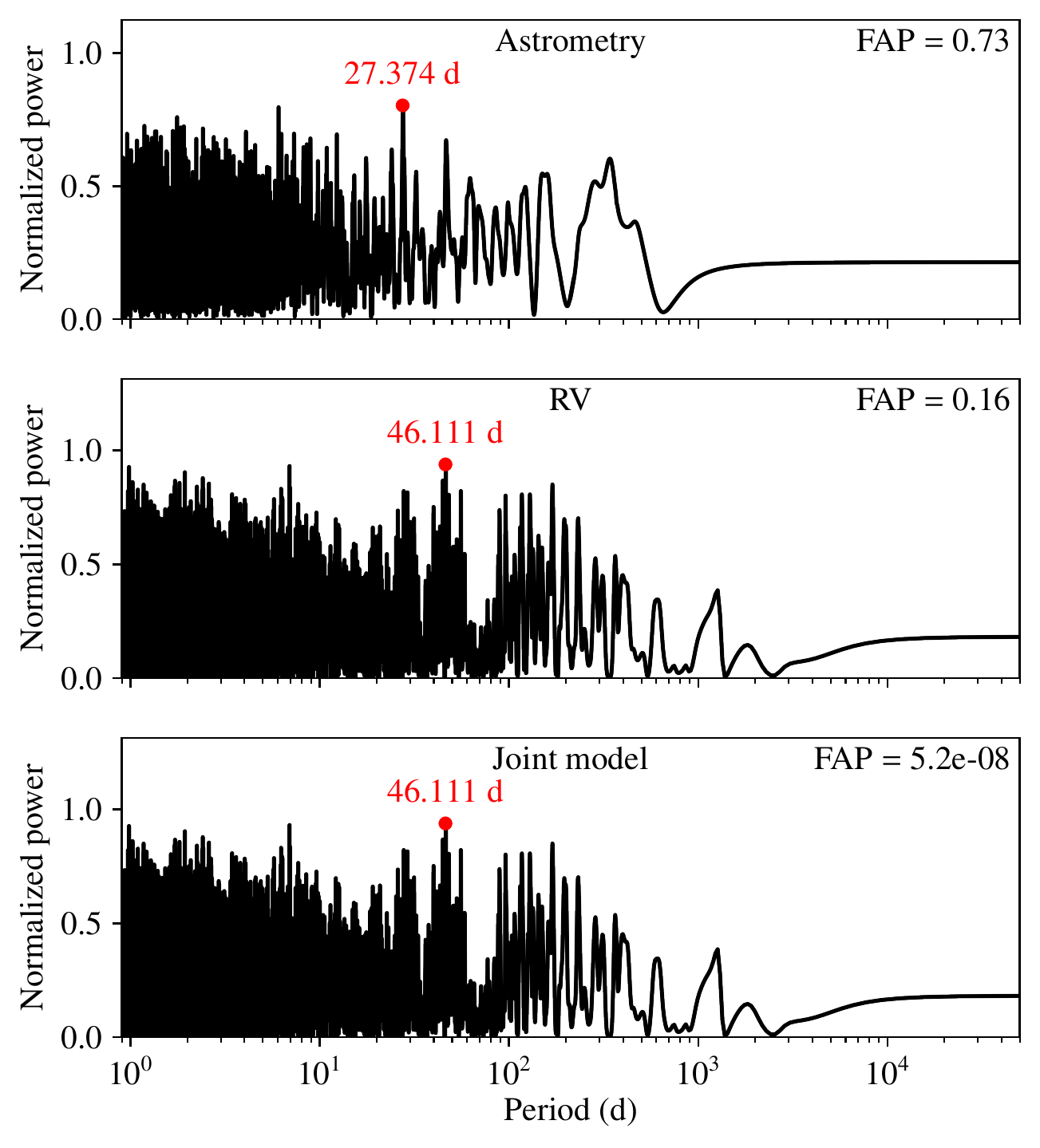}
  \caption{Periodograms of
    the Hipparcos time series (astrometry, \textit{top}),
    CORALIE time series (RV, \textit{middle}),
    and joint periodogram (\textit{bottom}) of HD~3277,
    after fitting for its companion.}
  \label{fig:hd3277_perio_0}
\end{figure}

\begin{table*}
  \caption{Fit parameters of HD~3277.}
  \resizebox{\textwidth}{!}{\setlength{\tabcolsep}{1mm}
    \begin{tabular}{lcccccc}
      \hline
      \hline
                                    & \multicolumn{3}{c}{RV} & \multicolumn{3}{c}{Joint model}                                                                                                                                            \\
                                    & guess                  & fit                                  & fit incl. jitter                     & guess          & fit                                  & fit incl. jitter                     \\
      \hline
      $\pi$ (mas)                   & --                     & --                                   & --                                   & $       34.26$ & $       34.37 \pm         0.50$      & $       34.38 \pm         0.56$      \\
      $\Delta \delta$ (mas)         & --                     & --                                   & --                                   & $      -750.5$ & $      -752.6 \pm          1.9$      & $      -752.6 \pm          2.1$      \\
      $\Delta \alpha^*$ (mas)       & --                     & --                                   & --                                   & $       525.9$ & $       522.2 \pm          2.1$      & $       522.3 \pm          2.3$      \\
      $\mu_\delta$ (mas/yr)         & --                     & --                                   & --                                   & $     -165.99$ & $     -166.52 \pm         0.42$      & $     -166.53 \pm         0.45$      \\
      $\mu_\alpha$ (mas/yr)         & --                     & --                                   & --                                   & $      116.29$ & $      115.73 \pm         0.47$      & $      115.76 \pm         0.52$      \\
      $\sigma_\mathrm{HIP}$ (mas)   & --                     & --                                   & --                                   & --             & --                                   & $        0.50 \pm         0.38$      \\
      \hline
      $\gamma_\mathrm{COR98}$ (m/s) & $    -14272.8$         & $    -14691.8 \pm          4.2$      & $    -14687.6 \pm          4.2$      & $    -14272.8$ & $    -14691.8 \pm          4.2$      & $    -14687.6 \pm          4.2$      \\
      $\gamma_\mathrm{COR07}$ (m/s) & $    -14837.0$         & $    -14693.5 \pm          4.7$      & $    -14694.9 \pm          2.3$      & $    -14837.0$ & $    -14693.5 \pm          4.7$      & $    -14694.9 \pm          2.3$      \\
      $\sigma_\mathrm{COR98}$ (m/s) & --                     & --                                   & $         3.7 \pm          3.2$      & --             & --                                   & $         3.7 \pm          3.2$      \\
      $\sigma_\mathrm{COR07}$ (m/s) & --                     & --                                   & $         0.0 \pm          2.0$      & --             & --                                   & $         0.0 \pm          2.0$      \\
      \hline
      $P$ (d)                       & $    46.11119$         & $    46.15131 \pm 2.2\times 10^{-4}$ & $    46.15116 \pm 1.6\times 10^{-4}$ & $    46.11119$ & $    46.15131 \pm 2.2\times 10^{-4}$ & $    46.15116 \pm 1.6\times 10^{-4}$ \\
      $M_0+\omega$ (deg)            & $       44.16$         & $       70.50 \pm         0.15$      & $       70.37 \pm         0.13$      & $       44.16$ & $       70.49 \pm         0.15$      & $       70.37 \pm         0.13$      \\
      $a_\mathrm{s} \sin i$ (AU)    & $   0.0172900$         & $    0.016563 \pm 1.8\times 10^{-5}$ & $   0.0165860 \pm 8.2\times 10^{-6}$ & $   0.0172900$ & $    0.016563 \pm 1.8\times 10^{-5}$ & $   0.0165860 \pm 8.2\times 10^{-6}$ \\
      $e$                           & $     0.21570$         & $     0.28532 \pm 9.4\times 10^{-4}$ & $     0.28437 \pm 6.3\times 10^{-4}$ & $     0.21570$ & $     0.28532 \pm 9.4\times 10^{-4}$ & $     0.28437 \pm 6.3\times 10^{-4}$ \\
      $\omega$ (deg)                & $      350.76$         & $      320.55 \pm         0.24$      & $      320.69 \pm         0.18$      & $      350.76$ & $      320.56 \pm         0.24$      & $      320.69 \pm         0.18$      \\
      $i$ (deg)                     & --                     & --                                   & --                                   & $       126.8$ & $       167.5 \pm          2.0$      & $       167.5 \pm          2.2$      \\
      $\Omega$ (deg)                & --                     & --                                   & --                                   & $       223.5$ & $       264.8 \pm          9.0$      & $       265.5 \pm          9.9$      \\
      \hline
      $\log\mathcal{L}$             & -13101.91              & -48.98                               & -46.64                               & -13152.82      & -87.78                               & -85.13                               \\
      \hline
    \end{tabular}
  }
  \tablefoot{The reference epoch is 2\,450\,000~BJD.}
  \label{tab:hd3277}
\end{table*}

\begin{figure}
  \centering
  \includegraphics[width=0.7\linewidth]{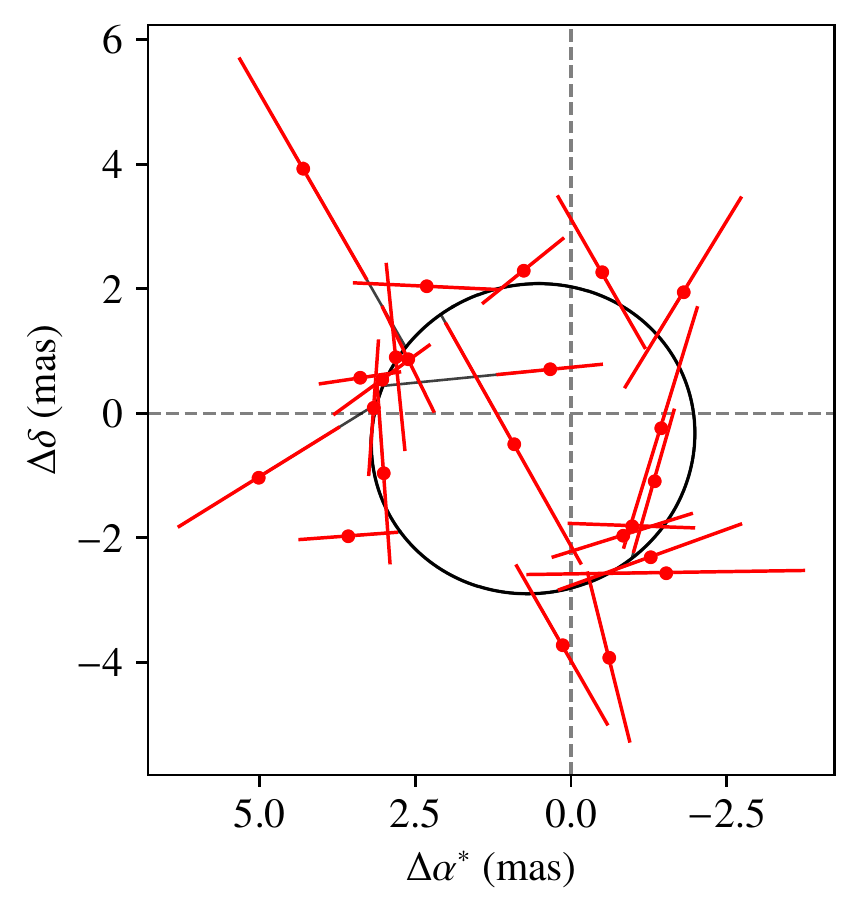}
  \caption{Astrometric orbit of HD~3277.}
  \label{fig:hd3277_astro_orbit}
\end{figure}

\begin{figure}
  \centering
  \includegraphics[width=\linewidth]{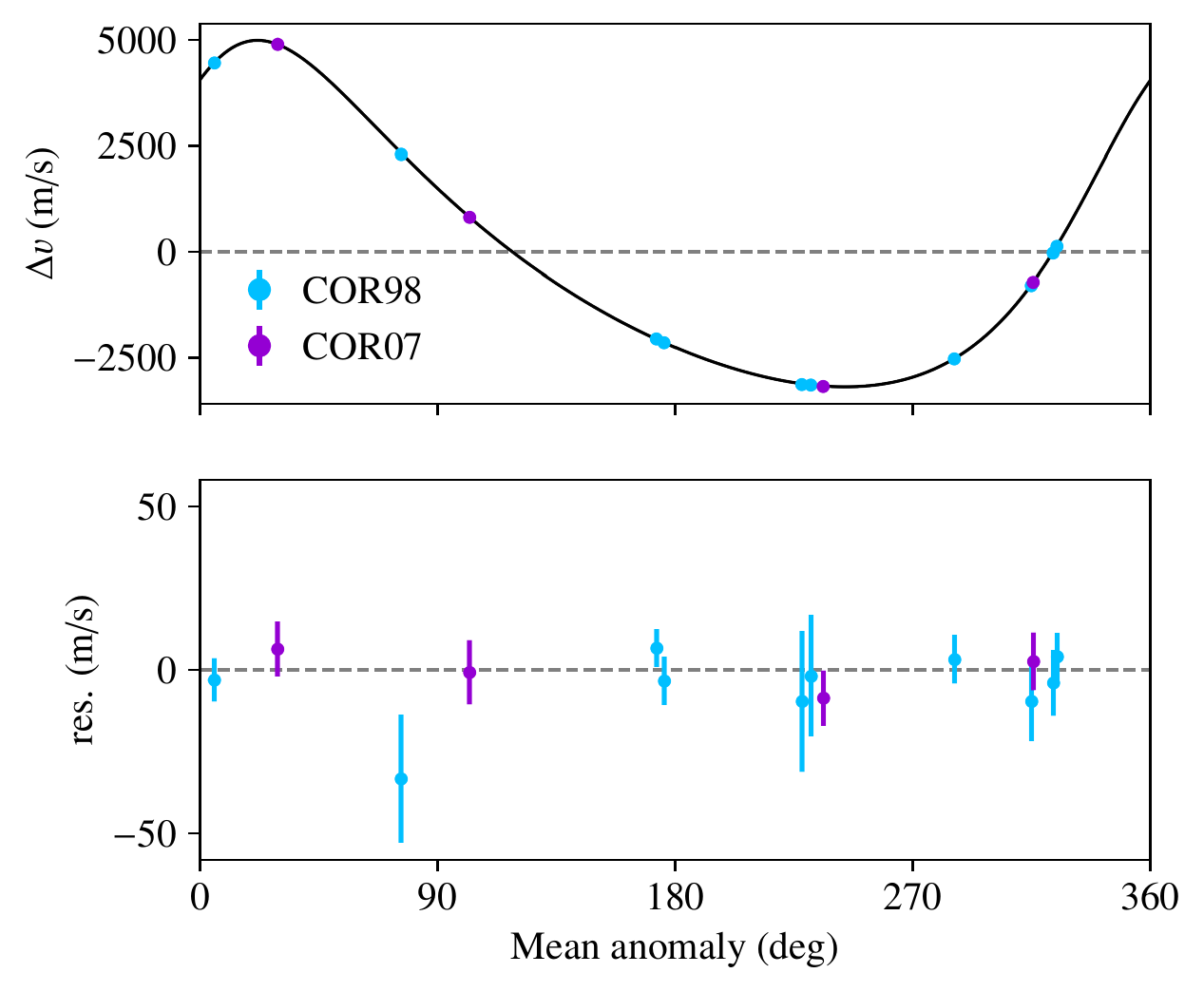}
  \caption{Phase-folded RV time series of HD~3277.}
  \label{fig:hd3277_rv_orbit}
\end{figure}

\begin{figure}
  \centering
  \includegraphics[width=\linewidth]{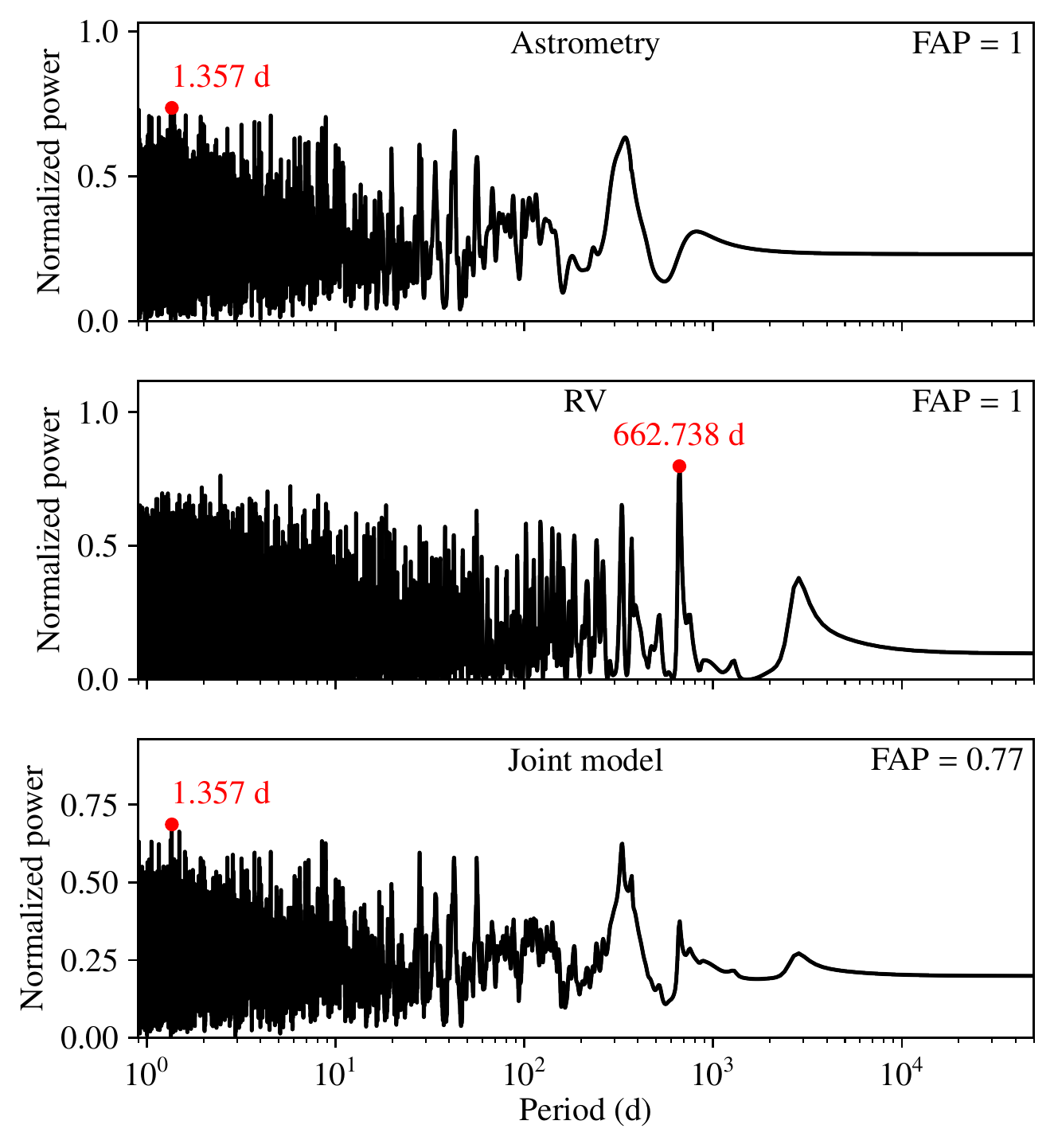}
  \caption{Periodograms of the residuals of
    the Hipparcos time series (astrometry, \textit{top}),
    CORALIE time series (RV, \textit{middle}),
    and joint periodogram (\textit{bottom}) of HD~3277,
    after fitting for its companion.}
  \label{fig:hd3277_perio_1}
\end{figure}

The last system we analyzed is HD~3277 (HIP~2790).
It was first listed by \citet{tokovinin_2006_tertiary} as an uncertain spectroscopic
binary based on few CORALIE measurements
and \citet{sahlmann_2011_search} confirmed the presence of a companion
using additional CORALIE measurements as well as Hipparcos data.
The primary component is a G8 star while the companion is an M-dwarf.
We analyzed the same data as in \citet{sahlmann_2011_search}, namely,
Hipparcos data and 14 CORALIE measurements (10 COR98 and 4 COR07).

The periodograms and FAP of the astrometric and radial velocity time series
are shown in Fig.~\ref{fig:hd3277_perio_0}.
We observe a non-significant peak around 46~d in the radial velocities periodogram
($\mathrm{FAP} = 0.16$).
This high FAP value can be explained by the poor sampling of the radial velocity time series,
which consists of only 14 points split among two instruments.
To test this hypothesis, we recomputed the periodogram and FAP of the radial velocity time series
but assuming the offset between COR98 and COR07 to be negligible,
and found a more significant FAP of 0.015.
A peak is also visible in the astrometric periodogram at the same period,
but it is not the highest one, and the FAP is very high (0.73).
The joint periodogram is very similar in shape to the radial velocities periodogram
but the associated FAP is significantly lower ($5 \times 10^{-8}$).

Since the 46~d signal is not detected using astrometry alone, we only applied our procedure
to the radial velocities and the joint model.
The parameters we obtained are shown in Table~\ref{tab:hd3277}.
As for previous systems, our analytical method provides a good initial guess of the parameters,
which is then refined by maximizing the likelihood.
The most challenging parameter is the inclination, which is only constrained by the astrometry,
for which the initial guess is within 40~deg of the maximum likelihood value.
This slight discrepancy is not surprising considering the low S/N of the signal
in the astrometric time series (high FAP, other more significant peaks in the periodogram).

\section{Discussion}
\label{sec:discussion}

In this work, we present a set of methods
to detect and characterize companions (planets, brown dwarfs, and stars)
in astrometric or radial velocity time series, or both.
We propose a general linear periodogram framework with an analytical method to compute the associated FAP.
We additionally provide an analytical method to obtain an estimate of the orbital parameters
once the period of the companion has been determined (from the periodogram approach).
These methods can be applied to astrometric data alone, radial velocities alone,
but also when combining both detection techniques.

Being based on analytical approximations, our methods are very efficient in terms of computational cost.
This is of particular interest in the context of the Gaia space mission,
considering the amount of generated data.
It should be noted that these approximate analytical methods are complementary
with more precise numerical methods.
For instance, in Sect.~\ref{sec:applications}, we use our analytical estimate of the orbital parameters
as a starting point for a likelihood maximization algorithm.
More generally, both the periodogram-FAP approach and the orbital parameters approximation
can be used to improve the convergence speed of numerical algorithms,
such as Markov chain Monte-Carlo, genetic algorithms, and so on.
In particular, our analytical estimate of orbital parameters
has been implemented in the genetic algorithm of the official Gaia pipeline for the detection of planets
\citep[MIKS-GA, see][]{holl_2022_gaia}.

Indeed, a primary objective of this article is to provide building blocks for
the detection and characterization of planets using Gaia astrometry alone
as well as in combination with on-ground radial velocities.
Since the Gaia collaboration will not publish astrometric time series before DR4,
we were not able to illustrate our methods on Gaia sources hosting planetary companions.
We instead chose to reanalyze the Hipparcos astrometric time series of three stars
known to host a stellar companion:
\object{HD~223636} (HIP~117622), \object{HD~17289} (HIP~12726), and \object{HD~3277} (HIP~2790).
Using Hipparcos astrometry and publicly available CORALIE radial velocities,
we obtained results similar to the literature.
In these examples, we found that the joint periodograms (astrometry + radial velocities)
were very similar to the periodogram of the radial velocity alone.
This is because the S/N of the companion is much higher in CORALIE radial velocities
than in Hipparcos astrometry.
Therefore, a combined periodogram might not actually appear necessary to find the companion period
in these cases.
However, since Gaia will provide much higher precision astrometric time series
and with a much longer time span than Hipparcos,
the contribution of astrometry and radial velocities will be more balanced.

Finally, it should be noted that while we assume white noise in our illustrations (Sect.~\ref{sec:applications}),
our methods are derived in the more general case of correlated noise.
Indeed, radial velocities are known to be affected by stellar activity,
which is often modeled using Gaussian processes (i.e., correlated noise models).
Other sources of noise, such as the spectrograph calibration noise,
can also introduce correlated noise in the radial velocity time series \citep[e.g.,][]{delisle_2020_efficientb}.
In the case of astrometric data, the stellar activity signature
is expected to be much smaller than for radial velocities
\citep{lagrange_2011_using}.
However, the attitude calibration of the satellite
might introduce correlated noise in the astrometric time series.
In the case of Gaia, the correlated noise contribution remains to be investigated and characterized,
but once this is done, our framework is readily equipped to account for it.

\begin{acknowledgements}
  We thank the anonymous referee for their very constructive feedback that helped to improve this manuscript.
  We acknowledge financial support from the Swiss National Science Foundation (SNSF).
  This work has, in part, been carried out within the framework of
  the National Centre for Competence in Research PlanetS
  supported by SNSF.
\end{acknowledgements}

\bibliographystyle{aa}
\bibliography{draftastro}

\begin{appendix}

\section{Power fraction in the fundamental and harmonics}
\label{sec:power_fraction}

In this appendix, we compute the power fraction that is found in the fundamental
and in the harmonics for a radial velocity (Appendix~\ref{sec:power_rv})
and an astrometric Keplerian signal (Appendix~\ref{sec:power_astro}).

\subsection{Radial velocity signal}
\label{sec:power_rv}

We follow \citet{baluev_2015_keplerian} to define the total power of a Keplerian radial velocity signal $V_K(t) = K(\cos(v(t) + \omega) + e
  \cos\omega)$ as:
\begin{align}
  \mathcal{P}_\mathrm{tot.} & = \frac{1}{P}\int_0^P V_K^2(t) \d t\nonumber                           \\
                            & = \frac{K^2}{2} \left(1-e^2\right) \left(1-\beta^2\cos 2\omega\right),
\end{align}
where
\begin{equation}
  \beta = \frac{e}{1+\sqrt{1-e^2}},
\end{equation}
and $v$ is the true anomaly.
The Fourier expansion of this signal is
\citep[e.g.,][]{delisle_2016_analytical}:
\begin{align}
  V_0 & = 0,\nonumber                                                                                               \\
  V_k & = \frac{K\expo{\i k M_0}}{2}\left(X_k \expo{\i \omega} + X_{-k} \expo{-\i \omega}\right) \qquad (k \neq 0),
\end{align}
where $X_k$ is the Hansen coefficient $X_k^{0,1}$ which only depends on the eccentricity $e$
and is defined has
\begin{equation}
  X_k(e) = \frac{1}{2\pi} \int_0^{2\pi} \expo{\i v} \expo{-\i k M} \d M.
\end{equation}
The power in the harmonics are thus:
\begin{align}
  \mathcal{P}_0 & = 0,                                                                                      \\
  \mathcal{P}_k & = 2 V_k \bar{V}_k = \frac{K^2}{2} \left(X_k^2+X_{-k}^2 + 2 X_k X_{-k} \cos 2\omega\right)
  \qquad (k > 0).\nonumber
\end{align}
Finally the power fraction in the $k$-th harmonics ($k> 0$) is:
\begin{equation}
  \mathcal{R}_k = \frac{\mathcal{P}_k}{\mathcal{P}_\mathrm{tot.}} = \frac{X_k^2+X_{-k}^2 + 2 X_k X_{-k} \cos 2\omega}{\left(1-e^2\right) \left(1-\beta^2\cos 2\omega\right)}.
\end{equation}
As shown by \citet{baluev_2015_keplerian}, the ratio $\mathcal{R}_1$
is linked with the detection efficiency of the classical periodogram
(which only uses the fundamental frequency)
compared to a more complex Keplerian periodogram (which uses the full information contained in the signal).

The power fraction $\mathcal{R}_k$
depends on the eccentricity and (weakly) on the argument of periastron.
For a given eccentricity, the extrema of the power fraction correspond to $2\omega=0$ or $\pi$.
We define
\begin{align}
  \mathcal{R}_k^+ & = \frac{X_k^2+X_{-k}^2 + 2 X_k X_{-k}}{\left(1-e^2\right) \left(1-\beta^2\right)},\nonumber \\
  \mathcal{R}_k^- & = \frac{X_k^2+X_{-k}^2 - 2 X_k X_{-k}}{\left(1-e^2\right) \left(1+\beta^2\right)},
\end{align}
such that
$\mathcal{R}_k$ is always between $\mathcal{R}_k^+$ and $\mathcal{R}_k^-$ (the order depending on $k$ and $e$).
In Fig.~\ref{fig:power_fraction}, we plot the possible range of values of $\mathcal{R}_k$ as a function of
the eccentricity, for $k=1,2$ (fundamental and first harmonics).
For this plot, we evaluated the Hansen coefficients by numerical integration over the eccentric anomaly
\citep[see][Appendix A]{delisle_2016_analytical}.

\subsection{Astrometric signal}
\label{sec:power_astro}

Here, we follow the same approach as in the radial velocity case.
Since the astrometric signal has two dimensions, we define the total power
of the signal as the sum of powers along both directions:
\begin{align}
  \mathcal{P}_\mathrm{tot.} & = \frac{1}{P}\int_0^P \left(\delta_K^2(t) + \alpha_K^2(t)\right) \d t\nonumber \\
                            & = \frac{1}{2\pi}\int_0^{2\pi} \left((Ax+Fy)^2 + (Bx+Gy)^2\right) \d M.
\end{align}
We have
\begin{align}
  \frac{1}{2\pi}\int_0^{2\pi} x^2 \d M & = \frac{1}{2\pi}\int_0^{2\pi} x^2 \left(1-e\cos E\right)\d E\nonumber                                      \\
                                       & = \frac{1}{2\pi}\int_0^{2\pi} \left(\cos E - e\right)^2 \left(1-e\cos E\right)\d E\nonumber                \\
                                       & = \frac{1}{2} + 2 e^2,                                                                                     \\
  \frac{1}{2\pi}\int_0^{2\pi} y^2 \d M & = \frac{1-e^2}{2\pi}\int_0^{2\pi} \sin^2 E \left(1-e\cos E\right)\d E\nonumber                             \\
                                       & = \frac{1-e^2}{2},                                                                                         \\
  \frac{1}{2\pi}\int_0^{2\pi} xy \d M  & = \frac{\sqrt{1-e^2}}{2\pi}\int_0^{2\pi} \left(\cos E - e\right)\sin E \left(1-e\cos E\right)\d E\nonumber \\
                                       & = 0,
\end{align}
thus:
\begin{align}
  \mathcal{P}_\mathrm{tot.} & = \left(\frac{1}{2} + 2 e^2\right)\left(A^2+B^2\right) + \frac{1-e^2}{2}\left(F^2+G^2\right)\nonumber \\
                            & = \frac{a_s^2}{4}\left(\left(2 + 3 e^2\right)\left(2-\sin^2i\right)
  + 5 e^2 \sin^2 i \cos 2\omega\right).
\end{align}

The Fourier expansion of $\delta_K$ and $\alpha_K$ is provided in Eqs.~(\ref{eq:defdkak}),~(\ref{eq:dkak}),
from which we deduce the power in the $k$-th harmonics ($k > 0$):
\begin{align}
  \mathcal{P}_k & = 2\delta_k\bar{\delta}_k + 2\alpha_k\bar{\alpha}_k\nonumber                                                                      \\
                & = \frac{A^2+B^2+F^2+G^2}{2}\left(\zeta_k^2+\zeta_{-k}^2\right) + \left(A^2+B^2-F^2-G^2\right)\zeta_k\zeta_{-k}\nonumber           \\
                & = \frac{a_s^2}{2}\left(\left(\zeta_k^2+\zeta_{-k}^2\right)\left(2-\sin^2i\right) + 2\zeta_k\zeta_{-k}\sin^2 i\cos 2\omega\right).
\end{align}
In contrast to the case of radial velocities,
the Keplerian astrometric signal is not centered,
which implies that the power of the constant part ($k=0$) is not zero, that is,
\begin{align}
  \mathcal{P}_0 & = \delta_0^2 + \alpha_0^2 = \zeta_0^2 \left(A^2+B^2\right)\nonumber       \\
                & = \frac{9}{8} a_s^2 e^2\left(2 - \sin^2 i + \sin^2 i \cos 2\omega\right).
\end{align}
In practice, the constant part does not contain any information about the companion
because the reference position of the primary always needs to be adjusted at the same time as the orbit.
Therefore, in the following, we normalize the power in the harmonics by the useful total power:
\begin{equation}
  \mathcal{P}_\mathrm{tot.}-\mathcal{P}_0 = \frac{a_s^2}{8}\left(\left(4 - 3 e^2\right)\left(2-\sin^2i\right)
  + e^2 \sin^2 i \cos 2\omega\right),
\end{equation}
which is equivalent to re-centering the Keplerian astrometric model.
We then define the power fraction as
\begin{align}
  \mathcal{R}_k & = \frac{\mathcal{P}_k}{\mathcal{P}_\mathrm{tot.}-\mathcal{P}_0}\nonumber                                                                                                                 \\
                & = 4 \frac{\left(\zeta_k^2+\zeta_{-k}^2\right)\left(2-\sin^2i\right) + 2\zeta_k\zeta_{-k}\sin^2 i\cos 2\omega}{\left(4 - 3 e^2\right)\left(2-\sin^2i\right) + e^2 \sin^2 i \cos 2\omega}.
\end{align}
This fraction depends on $e$, $\omega$ and $i$.
For a given eccentricity, the extrema of the fraction appears for $2\omega = 0, \pi$ and $i=\pi/2$.
We thus define
\begin{align}
  \mathcal{R}_k^+ & = 2 \frac{\left(\zeta_k+\zeta_{-k}\right)^2}{2 - e^2},\nonumber \\
  \mathcal{R}_k^- & = \frac{\left(\zeta_k-\zeta_{-k}\right)^2}{1 - e^2},
\end{align}
such that $\mathcal{R}_k$ is always between $\mathcal{R}_k^+$ and $\mathcal{R}_k^-$.
We then need to evaluate the coefficients $\zeta_k$ to compute these fractions.
We use the same method as for the Hansen coefficient (Appendix~\ref{sec:power_rv}).
We estimate the integral of Eq.~(\ref{eq:zetakEA}) numerically by sampling $E$ in $[0, 2\pi]$,
\begin{align}
  \zeta_k \approx \frac{1}{N} \sum_{j=0}^{N-1} & \left(\cos E_j - e + \i \sqrt{1-e^2}\sin E_j\right)\nonumber \\
                                               & \times \expo{-\i k (E_j-e\sin E_j)} (1-e\cos E_j),
\end{align}
with $E_j = 2\pi k/N$.
A very good precision is achieved using $N=100$ \citep[see][]{delisle_2016_analytical},
which is the value we use to compute the fractions shown in Fig.~\ref{fig:power_fraction}.

\section{Periodogram definitions and computation of the FAP}
\label{sec:fapcomputation}

In this appendix, we provide details on the computation of the periodogram FAP
in the case of astrometric data alone
and in the case of a combination of astrometric and radial velocity time series.
The FAP computation takes into account correlated noise.
Our approach is very similar to the one used in \citet{delisle_2020_efficient} for radial velocity time series.
It is an extension to the correlated noise case of the method of \citet{baluev_2008_assessing},
which assumed white noise.

In addition to GLS periodogram defined in Eq.~(\ref{eq:GLS}),
\citet{baluev_2008_assessing} proposed four alternative periodogram definitions:
\begin{align}
  \label{eq:defperio}
  z_0(\nu) = \frac{1}{2}\left(\chi^2_\H-\chi^2_\K(\nu)\right), \qquad                 &
  z_1(\nu) = \frac{n_\H}{2} \frac{\chi_\H^2 - \chi_\K^2(\nu)}{\chi_\H^2},\nonumber      \\
  z_2(\nu) = \frac{n_\K}{2} \frac{\chi_\H^2 - \chi_\K^2(\nu)}{\chi_\K^2(\nu)}, \qquad &
  z_3(\nu) = \frac{n_\K }{2} \ln \frac{\chi_\H^2}{\chi_\K^2(\nu)},
\end{align}
where $n_\H = n - p$ and $n_\K = n - (p + d)$.
The definitions $z_{1,2,3}$ are actually related to each other
and to the GLS periodogram.
Indeed, we have:
\begin{align}
  \label{eq:z123}
  z_1(\nu) & = \frac{n_\H}{2} z_\mathrm{GLS}(\nu),\nonumber                               \\
  z_2(\nu) & = \frac{n_\K}{2} \frac{z_\mathrm{GLS}(\nu)}{1-z_\mathrm{GLS}(\nu)},\nonumber \\
  z_3(\nu) & = - \frac{n_\K }{2} \ln (1-z_\mathrm{GLS}(\nu)).
\end{align}
In the following, we only consider the GLS (Eq.~(\ref{eq:GLS}))
and the definition $z_0$ (Eq.~(\ref{eq:defperio})) of the periodogram,
but all the results on the GLS also apply to $z_{1,2,3}$ (using Eq.~(\ref{eq:z123})).

\begin{table}
  \caption{False alarm probability for the two definitions (Eqs.~(\ref{eq:GLS}),~(\ref{eq:defperio}))
    of the periodogram power \citep[see][]{baluev_2008_assessing}.}
  {\renewcommand{\arraystretch}{2}\setlength{\tabcolsep}{3pt}\small\centering
    \begin{tabular}{ll|ll}
      \hline
      \hline
      $z$                                                                                                            & $d$ & $\mathrm{FAP_{single}}(Z)$ & $\tau(Z, \nu_\mathrm{max})$ \\
      \hline
      $z_\mathrm{GLS}$                                                                                               & --  &
      $\displaystyle 1-\frac{\mathrm{B}(Z; d/2, n_\K/2)}{\mathrm{B}(d/2, n_\K/2)}$                                   &
      $\gamma W \left(1-Z\right)^\frac{n_\K-1}{2} Z^{\frac{d-1}{2}}$                                                                                                                  \\
                                                                                                                     & 2   &
      $\left(1-Z\right)^\frac{n_\K}{2}$                                                                              &
      $\gamma W\left(1-Z\right)^\frac{n_\K-1}{2}\sqrt{Z}$                                                                                                                             \\
                                                                                                                     & 4   &
      $\left(1+\frac{n_\K}{2}Z\right)\left(1-Z\right)^\frac{n_\K}{2}$                                                &
      $\gamma W\left(1-Z\right)^\frac{n_\K-1}{2}Z^{\frac{3}{2}}$                                                                                                                      \\
                                                                                                                     & 6   &
      $\left(1+\frac{n_\K}{2}Z+\frac{n_\K}{2}\left(\frac{n_\K}{2}+1\right)Z^2\right)\left(1-Z\right)^\frac{n_\K}{2}$ &
      $\gamma W\left(1-Z\right)^\frac{n_\K-1}{2}Z^{\frac{5}{2}}$                                                                                                                      \\
      \hline
      $z_0$                                                                                                          & --  &
      $\displaystyle\frac{\Gamma(d/2, Z)}{\Gamma(d/2)}$                                                              &
      $W \expo{-Z} Z^{\frac{d-1}{2}}$                                                                                                                                                 \\
                                                                                                                     & 2   &
      $\expo{-Z}$                                                                                                    &
      $W\expo{-Z}\sqrt{Z}$                                                                                                                                                            \\
                                                                                                                     & 4   &
      $(1+Z)\expo{-Z}$                                                                                               &
      $W\expo{-Z}Z^{\frac{3}{2}}$                                                                                                                                                     \\
                                                                                                                     & 6   &
      $\left(1+Z+\frac{Z^2}{2}\right)\expo{-Z}$                                                                      &
      $W\expo{-Z}Z^{\frac{5}{2}}$                                                                                                                                                     \\
      \hline
    \end{tabular}}
  \tablefoot{The factor $W$ is the rescaled frequency bandwidth
  defined in Eq.~(\ref{eq:defWapp}),
  $\mathrm{B}(a, b)$ is the beta function, $\mathrm{B}(x; a, b)$ is the incomplete beta function,
  $\Gamma(s)$ is Euler's gamma function, and $\Gamma(s, x)$ is the upper incomplete gamma function.
  The factor
  $\gamma = \Gamma\left(\frac{n_\H}{2}\right)/\Gamma\left(\frac{n_\K+1}{2}\right)$
  can be approximated by $\gamma \sim \left(\frac{n_\H}{2}\right)^{\frac{d-1}{2}}$ for $n_\H \geq 10$. }
  \label{tab:powerfull}
\end{table}

The FAP can be bounded by \citep[see][Eq.~(5)]{baluev_2008_assessing}
\begin{equation}
  \mathrm{FAP_{max}}(Z, \nu_\mathrm{max}) \leq \mathrm{FAP_{single}}(Z) + \tau(Z, \nu_\mathrm{max}),
\end{equation}
and approximated by \citep[see][Eq.~(6)]{baluev_2008_assessing}
\begin{equation}
  \mathrm{FAP_{max}}(Z, \nu_\mathrm{max}) \approx 1 - \left(1-\mathrm{FAP_{single}}(Z)\right)\expo{-\tau(Z, \nu_\mathrm{max})},
\end{equation}
where $Z$ is the maximum periodogram power,
$\mathrm{FAP_{single}}(Z)$ is the FAP in the case in which the angular frequency $\nu$
of the putative additional signal is fixed,
$\tau(Z, \nu_\mathrm{max})$ is the expectation of the number of up-crossings of the level $Z$
by the periodogram \citep[see][]{baluev_2008_assessing}.
Computing $\mathrm{FAP_{single}}(Z)$ and $\tau(Z, \nu_\mathrm{max})$
requires us to specify the definition of the periodogram $z(\nu)$.
\citet{baluev_2008_assessing} proposed several definitions
and derived the corresponding formulas for $\mathrm{FAP_{single}}(Z)$ and $\tau(Z, \nu_\mathrm{max})$.
These results are summarized in Table~\ref{tab:powerfull},
for the definitions of the periodogram of Eqs.~(\ref{eq:GLS}),~(\ref{eq:defperio}).
The only quantity left to compute is the factor $W$,
which is the rescaled frequency bandwidth, defined as \citep[see][]{baluev_2008_assessing}
\begin{equation}
  \label{eq:defWapp}
  W = \frac{A(\nu_\mathrm{max})}{2\pi^{(d+1)/2}},
\end{equation}
where
\begin{equation}
  \label{eq:Anumax}
  A(\nu_\mathrm{max}) = \int_{0}^{\nu_\mathrm{max}}\d\nu \int_{x^2<1} \sqrt{\frac{x\t M(\nu) x}{(x\t x)^d}} \d x.
\end{equation}
The $d\times d$ matrix $M(\nu)$ is defined as follows \citep[see][]{delisle_2020_efficient}
\begin{align}
  \label{eq:defM}
   & Q = \varphi\t C^{-1} \varphi, \qquad
  S = \varphi\t C^{-1} \varphi',\nonumber                                           \\
   & R = \varphi\primet C^{-1} \varphi',\nonumber                                   \\
   & Q_\H = \varphi_\H\t C^{-1} \varphi, \qquad
  S_\H =  \varphi_\H\t C^{-1} \varphi',\nonumber                                    \\
   & Q_{\H,\H} = \varphi_\H\t C^{-1} \varphi_\H,\nonumber                           \\
   & \tilde{Q} = Q - Q_\H\t Q_{\H,\H}^{-1} Q_\H, \qquad
  \tilde{S} = S - Q_\H\t Q_{\H,\H}^{-1} S_\H,\nonumber                              \\
   & \tilde{R} = R - S_\H\t Q_{\H,\H}^{-1} S_\H,\nonumber                           \\
   & M = \tilde{Q}^{-1}\left(\tilde{R} - \tilde{S}\t\tilde{Q}^{-1}\tilde{S}\right),
\end{align}
where $x' = \partial x/\partial\nu$.
\citet{baluev_2008_assessing} also defined the effective time series length as
\begin{equation}
  \label{eq:defTeff}
  T_\mathrm{eff} = \frac{A(\nu_\mathrm{max})}{\pi^{(d-1)/2}\nu_\mathrm{max}},
\end{equation}
such that
\begin{equation}
  W = \frac{\nu_\mathrm{max}}{2\pi} T_\mathrm{eff}.
\end{equation}
From Eqs.~(\ref{eq:Anumax}) and (\ref{eq:defTeff}), we obtain
\begin{equation}
  \label{eq:Teff}
  T_\mathrm{eff} = \frac{\pi^{(1-d)/2}}{\nu_\mathrm{max}}\int_0^{\nu_\mathrm{max}}\d\nu
  \int_{x^2<1} \sqrt{\frac{x\t M(\nu) x}{(x\t x)^d}} \d x.
\end{equation}
In the following we derive analytical approximations for $T_\mathrm{eff}$
in the case of astrometric data alone (Appendix~\ref{sec:fapd4})
and in the case of astrometric and radial velocity data (Appendix~\ref{sec:fapd6}).
The case of radial velocities alone was already treated in \citet{delisle_2020_efficient}.

\subsection{Astrometric time series}
\label{sec:fapd4}

In the case of astrometric data alone,
we have $d=4$ and
\begin{equation}
  \varphi(\nu) = (\cos\theta\cos\nu t, \sin\theta\cos\nu t,
  \cos\theta\sin\nu t, \sin\theta\sin\nu t).
\end{equation}
We thus replace this expression
in the definitions of $Q$, $S$, and $R$ (Eq.~(\ref{eq:defM})).
As an example, for $Q_{1,1}$ we obtain
\begin{align}
  Q_{1,1} & = \frac{1}{8} \sum_{i,j} C_{i,j}^{-1}\big(
  \cos(\nu t_i-\nu t_j-\theta_i-\theta_j) + \cos(\nu t_i-\nu t_j-\theta_i+\theta_j)\nonumber             \\
          & + \cos(\nu t_i-\nu t_j+\theta_i-\theta_j) + \cos(\nu t_i-\nu t_j+\theta_i+\theta_j)\nonumber \\
          & + \cos(\nu t_i+\nu t_j-\theta_i-\theta_j) + \cos(\nu t_i+\nu t_j-\theta_i+\theta_j)\nonumber \\
          & + \cos(\nu t_i+\nu t_j+\theta_i-\theta_j) + \cos(\nu t_i+\nu t_j+\theta_i+\theta_j)\big).
\end{align}
We follow \citet{baluev_2008_assessing,delisle_2020_efficient} and neglect aliasing effects.
In this approximation all the terms in $Q$, $S$, and $R$ that contain a sine or cosine of
$\nu t_i + \nu t_j \pm \theta_i \pm \theta_j$
average out when summing over $i, j$.
Similarly, the terms containing a sine or cosine of $\theta_i + \theta_j \pm \nu t_i \pm \nu t_j$ average out.
Finally, the terms containing $\sin(\nu t_i - \nu t_j \pm (\theta_i-\theta_j))$ can also be neglected.
Indeed, in the low frequency limit ($|\nu t_i - \nu t_j \pm (\theta_i-\theta_j)| \ll 1$), the sine vanishes,
while in the high frequency limit ($|\nu t_i - \nu t_j \pm (\theta_i-\theta_j)| \gg 1$), the sine average out.
Using these approximations, we obtain
\begin{align}
  \label{eq:QSRnoalias}
  Q & \approx q_+ U_+ + q_- U_-,\nonumber \\
  S & \approx s_+ V_+ + s_- V_-,\nonumber \\
  R & \approx r_+ U_+ + r_- U_-,
\end{align}
with
\begin{align}
  \label{eq:qsrU}
  q_\pm                  & = \frac{1}{4}\varmean{\cos(\nu \Delta_t \pm \Delta_\theta)},\nonumber                 \\
  s_\pm                  & = \frac{1}{8}\varmean{\Sigma_t\hadprod\cos(\nu \Delta_t \pm \Delta_\theta)},\nonumber \\
  r_\pm                  & = \frac{1}{4}\varmean{\Pi_t\hadprod\cos(\nu \Delta_t \pm \Delta_\theta)},\nonumber    \\
  U_+                    & = \frac{1}{2}\left(\begin{array}{cccc}
                                                  1  & 0 & 0 & -1 \\
                                                  0  & 1 & 1 & 0  \\
                                                  0  & 1 & 1 & 0  \\
                                                  -1 & 0 & 0 & 1  \\
                                                \end{array}\right),
                         & V_+ = \frac{1}{2}\left(\begin{array}{cccc}
                                                      0  & 1  & 1  & 0 \\
                                                      -1 & 0  & 0  & 1 \\
                                                      -1 & 0  & 0  & 1 \\
                                                      0  & -1 & -1 & 0 \\
                                                    \end{array}\right),\nonumber                                   \\
  U_-                    & = \frac{1}{2}\left(\begin{array}{cccc}
                                                  1 & 0  & 0  & 1 \\
                                                  0 & 1  & -1 & 0 \\
                                                  0 & -1 & 1  & 0 \\
                                                  1 & 0  & 0  & 1 \\
                                                \end{array}\right),
                         & V_- = \frac{1}{2}\left(\begin{array}{cccc}
                                                      0  & -1 & 1 & 0  \\
                                                      1  & 0  & 0 & 1  \\
                                                      -1 & 0  & 0 & -1 \\
                                                      0  & -1 & 1 & 0  \\
                                                    \end{array}\right),\nonumber                                   \\
  \varmean{X}            & = \sum_{i,j} C^{-1}_{i,j} X_{i,j},\nonumber                                           \\
  \Sigma_{t\,(i,j)}      & = t_i + t_j,\nonumber                                                                 \\
  \Delta_{t\,(i,j)}      & = t_i - t_j,\nonumber                                                                 \\
  \Pi_{t\,(i,j)}         & = t_i t_j,\nonumber                                                                   \\
  \Delta_{\theta\,(i,j)} & = \theta_i - \theta_j,
\end{align}
and
$\,\hadprod\,$ denotes the Hadamard (or elementwise) product.

Following \citet{baluev_2008_assessing,delisle_2020_efficient},
we additionally assume that $\tilde{Q}\approx Q$, $\tilde{S}\approx S$, $\tilde{R}\approx R$, and
\begin{equation}
  \label{eq:Mortho}
  M(\nu) = Q^{-1}\left(R - S\t Q^{-1}S\right).
\end{equation}
Replacing Eq.~(\ref{eq:QSRnoalias}) in Eq.~(\ref{eq:Mortho}), we find
\begin{equation}
  M(\nu) \approx \lambda_+ U_+ + \lambda_- U_-,
\end{equation}
with
\begin{equation}
  \label{eq:deflambdapm}
  \lambda_\pm = \frac{r_\pm}{q_\pm} - \left(\frac{s_\pm}{q_\pm}\right)^2.
\end{equation}
The eigenvalues of $M$ are thus $\lambda_\pm$, both with a multiplicity of 2.

In order to determine the effective time series length $T_\mathrm{eff}$ (Eq.~(\ref{eq:Teff}))
we then need to evaluate the integral
\begin{equation}
  I_d(\nu) = \int_{x^2<1} \sqrt{\frac{x\t M(\nu) x}{(x\t x)^d}} \d x.
\end{equation}
As shown by \citet{baluev_2008_assessing},
for a matrix $M$ with eigenvalues $\lambda_i$ ($i=1,\dots,d$),
the integral $I$ is linked to the surface-area $\Pi_d$
of the $d$-dimensional ellipsoid with semi-axes $1/\sqrt{\lambda_i}$,
through
\begin{equation}
  I_d = \Pi_d \sqrt{\det M}.
\end{equation}
Moreover, since in our case the eigenvalues have even multiplicities,
the surface area can be expressed in terms of elementary functions \citep[see][]{tee_2005_surface}.
Following the method described in \citet{tee_2005_surface}, we obtain
\begin{equation}
  \Pi_4 = \frac{4\pi^2}{3} \frac{1}{\lambda_+ \lambda_-} \left(
  \sqrt{\lambda_+} + \sqrt{\lambda_-}
  - \frac{\sqrt{\lambda_+\lambda_-}}{\sqrt{\lambda_+} + \sqrt{\lambda_-}}\right),
\end{equation}
and
\begin{equation}
  I_4 = \frac{4\pi^2}{3} \left(
  \sqrt{\lambda_+} + \sqrt{\lambda_-}
  - \frac{\sqrt{\lambda_+\lambda_-}}{\sqrt{\lambda_+} + \sqrt{\lambda_-}}\right).
\end{equation}
The effective time series length is thus (see Eq.~(\ref{eq:Teff})):
\begin{align}
  T_\mathrm{eff} & = \frac{\pi^{-3/2}}{\nu_\mathrm{max}} \int_0^{\nu_\mathrm{max}} I_4(\nu)\d\nu\nonumber \\
                 & = \frac{4\sqrt{\pi}}{3\nu_\mathrm{max}} \int_0^{\nu_\mathrm{max}} \left(
  \sqrt{\lambda_+} + \sqrt{\lambda_-}
  - \frac{\sqrt{\lambda_+\lambda_-}}{\sqrt{\lambda_+} + \sqrt{\lambda_-}}\right)\d\nu.
\end{align}
Finally, following \citet{delisle_2020_efficient}, we approximate the integral over the angular frequency $\nu$
by replacing $q_\pm$, $s_\pm$, and $r_\pm$ by their average over the range $]0,\nu_\mathrm{max}]$.
We have:
\begin{align}
  \mean{q}_\pm & = \frac{1}{\nu_\mathrm{max}}\int_0^{\nu_\mathrm{max}} q_\pm(\nu) \d\nu\nonumber \\
               & = \frac{1}{4} \varmean{
    \cos\left(\frac{\nu_\mathrm{max}\Delta_t}{2} \pm \Delta_\theta\right)
  \hadprod \sinc\left(\frac{\nu_\mathrm{max}\Delta_t}{2}\right)},\nonumber                       \\
  \mean{s}_\pm & = \frac{1}{8} \varmean{
    \Sigma_t
    \hadprod \cos\left(\frac{\nu_\mathrm{max}\Delta_t}{2} \pm \Delta_\theta\right)
  \hadprod \sinc\left(\frac{\nu_\mathrm{max}\Delta_t}{2}\right)},\nonumber                       \\
  \mean{r}_\pm & = \frac{1}{4} \varmean{
    \Pi_t
    \hadprod \cos\left(\frac{\nu_\mathrm{max}\Delta_t}{2} \pm \Delta_\theta\right)
    \hadprod \sinc\left(\frac{\nu_\mathrm{max}\Delta_t}{2}\right)},
\end{align}
and we make the following approximations
\begin{align}
  \label{eq:meanlambdapm}
  \mean{\lambda}_\pm & \approx \frac{\mean{r}_\pm}{\mean{q}_\pm} - \left(\frac{\mean{s}_\pm}{\mean{q}_\pm}\right)^2,\nonumber \\
  T_\mathrm{eff}     & \approx \frac{4\sqrt{\pi}}{3} \left(
  \sqrt{\mean{\lambda}_+} + \sqrt{\mean{\lambda}_-}
  - \frac{\sqrt{\mean{\lambda}_+\mean{\lambda}_-}}{\sqrt{\mean{\lambda}_+} + \sqrt{\mean{\lambda}_-}}\right).
\end{align}

\subsection{Astrometric and radial velocity time series}
\label{sec:fapd6}

In the case of a periodogram taking into account both astrometric and radial velocity time series,
we have $d=6$ and:
\begin{equation}
  \varphi(\nu) = \begin{pmatrix}
    \varphi_\mathrm{a}(\nu) & 0                         \\
    0                       & \varphi_\mathrm{rv}(\nu),
  \end{pmatrix}
\end{equation}
with
\begin{align}
  \varphi_\mathrm{a}(\nu)  & =
  (\cos\theta\cos\nu t_\mathrm{a}, \sin\theta\cos\nu t_\mathrm{a},
  \cos\theta\sin\nu t_\mathrm{a}, \sin\theta\sin\nu t_\mathrm{a}),\nonumber \\
  \varphi_\mathrm{rv}(\nu) & =
  (\cos\nu t_\mathrm{rv}, \sin\nu t_\mathrm{rv}).
\end{align}
Assuming the covariance matrix, $C$, to also be block diagonal,
\begin{equation}
  C = \begin{pmatrix}
    C_\mathrm{a} & 0              \\
    0            & C_\mathrm{rv},
  \end{pmatrix},
\end{equation}
the contributions of the astrometry and radial velocities in the matrix $Q$, $S$, $R$, and $M$
are completely separated and these matrices are also block diagonal.
Neglecting aliasing effects as in \citet{baluev_2008_assessing,delisle_2020_efficient}, and Appendix~\ref{sec:fapd4},
we find
\begin{align}
  Q & \approx q_+ U'_+ + q_- U'_- + q_\mathrm{rv} U'_\mathrm{rv},\nonumber          \\
  S & \approx s_+ V'_+ + s_- V'_- + s_\mathrm{rv} V'_\mathrm{rv},\nonumber          \\
  R & \approx r_+ U'_+ + r_- U'_- + r_\mathrm{rv} U'_\mathrm{rv},\nonumber          \\
  M & \approx \lambda_+ U'_+ + \lambda_- U'_- + \lambda_\mathrm{rv} U'_\mathrm{rv},
\end{align}
where
\begin{align}
   & U'_\pm = \begin{pmatrix}
                U_\pm & 0 \\
                0     & 0
              \end{pmatrix},\qquad
   & V'_\pm = \begin{pmatrix}
                V_\pm & 0 \\
                0     & 0
              \end{pmatrix},\nonumber         \\
   & U'_\mathrm{rv} = \begin{pmatrix}
                        0 & 0             \\
                        0 & U_\mathrm{rv}
                      \end{pmatrix},\qquad
   & V'_\mathrm{rv} = \begin{pmatrix}
                        0 & 0             \\
                        0 & V_\mathrm{rv}
                      \end{pmatrix},\nonumber \\
   & U_\mathrm{rv} = \mathbb{1},\qquad
   & V_\mathrm{rv} = \begin{pmatrix}
                       0  & 1 \\
                       -1 & 0
                     \end{pmatrix},
\end{align}
$q_\pm$, $s_\pm$, $r_\pm$, and $\lambda_\pm$ are defined according to Eqs.~(\ref{eq:qsrU}),~(\ref{eq:deflambdapm}),
and \citep[see][]{delisle_2020_efficient}
\begin{align}
  q_\mathrm{rv}               & = \frac{1}{2}\varmean{\cos\nu \Delta_\mathrm{rv}},\nonumber                                          \\
  s_\mathrm{rv}               & = \frac{1}{4}\varmean{\Sigma_\mathrm{rv}\hadprod\cos\nu \Delta_\mathrm{rv}},\nonumber                \\
  r_\mathrm{rv}               & = \frac{1}{2}\varmean{\Pi_\mathrm{rv}\hadprod\cos\nu \Delta_\mathrm{rv}},\nonumber                   \\
  \lambda_\mathrm{rv}         & = \frac{r_\mathrm{rv}}{q_\mathrm{rv}} - \left(\frac{s_\mathrm{rv}}{q_\mathrm{rv}}\right)^2,\nonumber \\
  \varmean{X}                 & = \sum_{i,j} C_\mathrm{rv\,(i,j)}^{-1} X_{i,j},\nonumber                                             \\
  \Sigma_{\mathrm{rv}\,(i,j)} & = t_{\mathrm{rv}\, i} + t_{\mathrm{rv}\, j},\nonumber                                                \\
  \Delta_{\mathrm{rv}\,(i,j)} & = t_{\mathrm{rv}\, i} - t_{\mathrm{rv}\, j},\nonumber                                                \\
  \Pi_{\mathrm{rv}\,(i,j)}    & = t_{\mathrm{rv}\, i} \  t_{\mathrm{rv}\, j}.
\end{align}
The eigenvalues of $M$ are thus $\lambda_+$, $\lambda_-$, and $\lambda_\mathrm{rv}$,
each with multiplicity 2.
Following the same steps as in Appendix~\ref{sec:fapd4}, we compute the surface-area $\Pi_6$
of the 6D ellipsoid with semi-axes $1/\sqrt{\lambda_i}$
using the method of \citet{tee_2005_surface} and find:
\begin{align}
  \Pi_6 =\  & \frac{8\pi^3}{15} \frac{1}{\lambda_+ \lambda_- \lambda_\mathrm{rv}} \left(
  \bracktwolines{
    \sqrt{\lambda_+} + \sqrt{\lambda_-} + \sqrt{\lambda_\mathrm{rv}}
  }{
    - \frac{
      \left(\sqrt{\lambda_+\lambda_-}
      + \sqrt{\lambda_+\lambda_\mathrm{rv}}
      + \sqrt{\lambda_-\lambda_\mathrm{rv}}\right)^2
    }{
      \left(\sqrt{\lambda_+} + \sqrt{\lambda_-}\right)
      \left(\sqrt{\lambda_+} + \sqrt{\lambda_\mathrm{rv}}\right)
      \left(\sqrt{\lambda_-} + \sqrt{\lambda_\mathrm{rv}}\right)
    }}\right).
\end{align}
We deduce
\begin{align}
  I_6 =\  & \Pi_6 \sqrt{\det M}\nonumber \\
  =\      & \frac{8\pi^3}{15} \left(
  \bracktwolines{
    \sqrt{\lambda_+} + \sqrt{\lambda_-} + \sqrt{\lambda_\mathrm{rv}}
  }{
    - \frac{
      \left(\sqrt{\lambda_+\lambda_-}
      + \sqrt{\lambda_+\lambda_\mathrm{rv}}
      + \sqrt{\lambda_-\lambda_\mathrm{rv}}\right)^2
    }{
      \left(\sqrt{\lambda_+} + \sqrt{\lambda_-}\right)
      \left(\sqrt{\lambda_+} + \sqrt{\lambda_\mathrm{rv}}\right)
      \left(\sqrt{\lambda_-} + \sqrt{\lambda_\mathrm{rv}}\right)
    }}\right),
\end{align}
and approximate the effective time series length as
\begin{align}
  T_\mathrm{eff} \approx\  & \frac{8\sqrt{\pi}}{15} \left(
  \bracktwolines{
    \sqrt{\mean\lambda_+} + \sqrt{\mean\lambda_-} + \sqrt{\mean\lambda_\mathrm{rv}}
  }{
    - \frac{
      \left(\sqrt{\mean\lambda_+\mean\lambda_-}
      + \sqrt{\mean\lambda_+\mean\lambda_\mathrm{rv}}
      + \sqrt{\mean\lambda_-\mean\lambda_\mathrm{rv}}\right)^2
    }{
      \left(\sqrt{\mean\lambda_+} + \sqrt{\mean\lambda_-}\right)
      \left(\sqrt{\mean\lambda_+} + \sqrt{\mean\lambda_\mathrm{rv}}\right)
      \left(\sqrt{\mean\lambda_-} + \sqrt{\mean\lambda_\mathrm{rv}}\right)
    }}\right),
\end{align}
with
$\mean\lambda_\pm$ defined as in Eq.~(\ref{eq:meanlambdapm})
and
\begin{align}
  \mean{\lambda}_\mathrm{rv} & \approx \frac{\mean{r}_\mathrm{rv}}{\mean{q}_\mathrm{rv}}
  - \left(\frac{\mean{s}_\mathrm{rv}}{\mean{q}_\mathrm{rv}}\right)^2,\nonumber                                                       \\
  \mean{q}_\mathrm{rv}       & = \frac{1}{2} \varmean{\sinc\nu_\mathrm{max}\Delta_\mathrm{rv}},\nonumber                             \\
  \mean{s}_\mathrm{rv}       & = \frac{1}{4} \varmean{\Sigma_\mathrm{rv} \hadprod \sinc\nu_\mathrm{max}\Delta_\mathrm{rv}},\nonumber \\
  \mean{r}_\mathrm{rv}       & = \frac{1}{2} \varmean{\Pi_\mathrm{rv} \hadprod \sinc\nu_\mathrm{max}\Delta_\mathrm{rv}}.
\end{align}

\section{Error propagation and weighted average for the eccentricity and the mean anomaly}
\label{sec:properror}

In this appendix, we provide details on the error propagation method used to
combine the estimates of the eccentricity
and mean anomaly at reference epoch obtained in Sect.~\ref{sec:guessastro} (astrometry alone)
or Sect.~\ref{sec:guesscomb} (astrometry and radial velocities).
We actually rather combine the estimates of $f = e \cos M_0$ and $g = e \sin M_0$.
For each estimate, we first linearize the problem in the form
\begin{equation}
  \d \hat{h}_k = \frac{\partial \hat{h}_k}{\partial \beta}\d\beta,
\end{equation}
where $h=f, g$ and $k = \delta, \alpha$ (astrometry alone) or $k = \delta, \alpha, \mathrm{rv}$ (astrometry and radial velocities).
The vector $\beta$ is the complete set of parameters obtained after a linear fit
($\beta=\beta_\mathrm{a}$ or $\beta = (\beta_\mathrm{a}, \beta_\mathrm{rv})$ depending on the considered case).
Then, we propagate the covariance matrix $C_\beta$ of the linear parameters
to approximate the covariance matrix $C_{\hat{\vec{h}}}$ of $\hat{\vec{h}}$
\begin{equation}
  C_{\hat{\vec{h}}} \approx \left(\frac{\partial \hat{\vec{h}}}{\partial \beta}\right)\t C_\beta \left(\frac{\partial \hat{\vec{h}}}{\partial \beta}\right).
\end{equation}
Finally, we compute the weighted average as
\begin{align}
  \hat{h}            & = \sigma_{\hat{h}}^2 u\t C_{\hat{\vec{h}}}^{-1} \hat{\vec{h}},\nonumber \\
  \sigma_{\hat{h}}^2 & = u\t C_{\hat{\vec{h}}}^{-1} u,
\end{align}
with $u = (1,1)$, or $(1,1,1)$, depending on the considered case.

The partial derivatives $\frac{\partial \hat{h}_k}{\partial \beta}$ needed to apply this method
can be derived from Eqs.~(\ref{eq:ratio})-(\ref{eq:M0da}).
We have
\begin{align}
  \d\hat{f}_k & = \cos\hat{M}_{0\,k} \d\hat{e}_k - \hat{g}_k \d\hat{M}_{0\,k},\nonumber \\
  \d\hat{g}_k & = \sin\hat{M}_{0\,k} \d\hat{e}_k + \hat{f}_k \d\hat{M}_{0\,k},
\end{align}
with
\begin{align}
  \d\hat{e}_k =\  & \frac{\d|\rho_k| + \hat{e}_k^3\d r_k}{1 - 3 r_k \hat{e}_k^2},\nonumber                                                    \\
  \d|\rho_k|  =\  & \left(\frac{\beta_{k,2,c}\d\beta_{k,2,c}+\beta_{k,2,s}\d\beta_{k,2,s}}{\beta_{k,2,c}^2 + \beta_{k,2,s}^2}\right.\nonumber \\
                  & \left.-\frac{\beta_{k,1,c}\d\beta_{k,1,c}+\beta_{k,1,s}\d\beta_{k,1,s}}{\beta_{k,1,c}^2 + \beta_{k,1,s}^2}
  \right)|\rho_k|,\nonumber                                                                                                                   \\
  \d r_k      =\  & \frac{1}{12}\sin(2\phi_k) \d\phi_k,\nonumber                                                                              \\
  \d\phi_k    =\  & 2\frac{\beta_{k,1,s}\d\beta_{k,1,c}-\beta_{k,1,c}\d\beta_{k,1,s}}{\beta_{k,1,c}^2 + \beta_{k,1,s}^2}\nonumber             \\
                  & -\frac{\beta_{k,2,s}\d\beta_{k,2,c}-\beta_{k,2,c}\d\beta_{k,2,s}}{\beta_{k,2,c}^2 + \beta_{k,2,s}^2}.
\end{align}
Moreover, we have (see Eq.~(\ref{eq:M0da}))
\begin{equation}
  \hat{M}_{0\,k} = \arg(\rho_k) - \arctan(b_k, a_k),
\end{equation}
with
\begin{align}
  a_k & = 1-r_k \hat{e}_k^2,\nonumber                \\
  b_k & = \frac{\hat{e}_k^2}{24}\sin(2\hat{\phi}_k).
\end{align}
We thus derive:
\begin{align}
  \d\hat{M}_{0\,k} =\  & \frac{\beta_{k,2,s}\d\beta_{k,2,c}-\beta_{k,2,c}\d\beta_{k,2,s}}{\beta_{k,2,c}^2 + \beta_{k,2,s}^2}\nonumber     \\
                       & -\frac{\beta_{k,1,s}\d\beta_{k,1,c}-\beta_{k,1,c}\d\beta_{k,1,s}}{\beta_{k,1,c}^2 + \beta_{k,1,s}^2}\nonumber    \\
                       & - \frac{a_k\d b_k - b_k\d a_k}{a_k^2 + b_k^2},\nonumber                                                          \\
  \d a_k          =\   & -\hat{e}_k^2 \d r_k - 2 r_k \hat{e}_k \d\hat{e}_k,\nonumber                                                      \\
  \d b_k          =\   & \frac{1}{12}\left(\hat{e}_k\sin(2\hat{\phi}_k)\d\hat{e}_k + \hat{e}_k^2\cos(2\hat{\phi}_k)\d\hat{\phi}_k\right).
\end{align}
\end{appendix}
\end{document}